\documentclass{aastex62}

\usepackage{hyperref}
\usepackage{graphicx}
\usepackage{rotating}
\usepackage{subfigure}
\setcitestyle{citesep={;}}

\received{\today}

\shorttitle{Neptune viewed with ALMA}
\shortauthors{Tollefson et al.}

\begin{document}

\title{Neptune's Latitudinal Variations as Viewed with ALMA}

\correspondingauthor{Joshua Tollefson}
\email{jtollefs@berkeley.edu}

\author{Joshua Tollefson}
\affil{University of California \\
Berkeley CA 94720, USA}

\author{Imke de Pater}
\affiliation{University of California \\
Berkeley CA 94720, USA}

\author{Statia Luszcz-Cook}
\affiliation{American Museum of Natural History \\
Columbia University \\
New York NY, USA}

\author{David DeBoer}
\affiliation{University of California \\
Berkeley CA 94720, USA}



\begin{abstract}
We present spatially resolved millimeter maps of Neptune between 95 and 242 GHz taken with the Atacama Large Millimeter/submillimeter Array (ALMA) in $2016-2017$. The millimeter weighting functions peak between 1 and 10 bar on Neptune, lying in between the altitudes probed at visible/infrared and centimeter wavelengths. Thus, these observations provide important constraints on the atmospheric structure and dynamics of Neptune.

We identify seven well-resolved latitudinal bands of discrete brightness temperature variations, on the order of $0.5-3$K in all three observed ALMA spectral bands. We model Neptune's brightness temperature using the radiative transfer code Radio-BEAR and compare how various H$_2$S, CH$_4$, and \textit{ortho/para} H$_2$ abundance profiles can fit the observed temperature variations across the disk. 
We find that observed variations in brightness temperature with latitude can be explained by variations in the H$_2$S profile that range from sub- to super-saturations at altitudes above the 10-bar pressure level, while variations in CH$_4$ improve the quality of fit near the equator. At the south polar cap, our best fit model has a depleted deep atmospheric abundance of H$_2$S from 30 to only 1.5 times the protosolar value, while simultaneously depleting the CH$_4$ abundance. This pattern of enhancement and depletion of condensable species is consistent with a global circulation structure where enriched air rises at the mid-latitudes ($32^{\circ}-12^{\circ}$S) and north of the equator ($2^{\circ}-20^{\circ}$N), and dry air descends at the poles ($90^{\circ}-66^{\circ}$S) and just south of the equator ($12^{\circ}$S$-2^{\circ}$N). Our analysis finds more complex structure near the equator than accounted for in previous circulation models.

\end{abstract}



\section{Introduction}
\label{S:1}

Millimeter continuum observations of Neptune provide a valuable bridge between visible/infrared studies of the cloud tops and above ($P<1$ bar) and deeply probing centimeter maps ($10<P<40$ bar). Visible and near-infrared (NIR) imaging from \textit{Voyager}, the Hubble Space Telescope (HST), and Keck have shown bright methane cloud activity at Neptune's mid-latitudes and comparatively dark regions and/or hazes at the equator and poles \citep{Limaye1991, Sromovsky2001b, Martin2012, Fitzpatrick2013, Tollefson2018}. Mid-infrared (MIR) images and spectra have been used to obtain zonal-mean temperature profiles which show that the equator and south pole are warmer than the southern mid-latitudes in the upper troposphere and stratosphere \citep{Conrath1991,Orton2007,Fletcher2014}. Centimeter maps obtained with the Very Large Array (VLA) probe well below Neptune's cloud deck and indicate the presence of dry (low-opacity) air at the south pole (e.g., \citealp{Butler2012DPS, dePater2014}). Taken together, observations of the deep troposphere and upper atmosphere are broadly consistent with a global circulation pattern where air enriched with trace gases rises at the mid-latitudes, adiabatically cools and condenses to form clouds, dries out, and descends at the poles and equator \citep{Conrath1991, dePater2014}. Millimeter observations probe altitudes in between those seen in the above maps, $1 < P < 10$ bar, making such observations vital for bridging the pressures viewed in the visible/infrared and centimeter, and for improving our overall understanding of Neptune's atmospheric structure and dynamics. In particular, millimeter continuum observations are sensitive to variations in composition, including variations in trace condensable species such as H$_2$S and CH$_4$. Like the centimeter, millimeter observed variations in these trace gases reflect atmospheric motions; depleted, low-opacity regions (enriched, high-opacity) are consistent with downwelling (upwelling) air. 

The opacity of Neptune's continuum at millimeter observations is dominated by collision-induced absorption of H$_2$ (CIA) with hydrogen, helium, and methane. Trace gases, such as H$_2$S, PH$_3$, and NH$_3$ also contribute to the overall opacity. Since millimeter observations of these gases probe Neptune's troposphere between 1 and 10 bar, individual lines are highly pressure-broadened, appearing as broad `continuum' bands in the millimeter spectrum. As a result, it is difficult to differentiate between these opacity sources since clear line features cannot be detected. 

On previous millimeter maps, Neptune's disk was spatially resolved and regions of enhanced or diminished brightness temperatures could be distinguished. Using the Combined Array for Research in Millimeter-wave Astronomy (CARMA), \citet{LuszczCook2013b} imaged Neptune in the far wings of the CO ($2-1$) line (230.538 GHz). In their longitudinally-averaged maps, they found brightness temperatures increased by $2-3$ K from $40^{\circ}$N to the south pole. Since their observations were taken far ($\sim 5$ GHz) from the CO ($2-1$) line center, this emission is primarily from sources forming the 'quasi-continuum'. Assuming an adiabat in the troposphere, the authors showed that variations in the brightness temperature at the south pole could be explained by a 30\% decrease in opacity at $P > 1$ bar. If variations in opacity would occur at altitudes only below 4 bar, the opacity needed to be decreased by a factor of 50. Hence, brightness variations at a particular wavelength are coupled to the pressure at which the opacity changes, itself dependent on the opacity source. \citet{LuszczCook2013b} could explain the latitudinal variations in brightness temperature as latitudinal gradients in any or all of: H$_2$S, CH$_4$, and/or \textit{ortho/para} H$_2$. However, due to their limited wavelength coverage, they were unable to disentangle the true contribution to the brightness from each candidate source.

In this paper, we present millimeter maps of Neptune taken with the Atacama Large Millimeter/Submillimeter Array (ALMA). ALMA provides the wavelength coverage, sensitivity, and resolution needed to constrain Neptune's opacity sources accurately across the disk. In Section 2, we present ALMA observations taken in three bands over the millimeter wavelength range ($1-3$ mm). The observed brightness temperature distribution is compared to model maps produced with the radiative-transfer code Radio-BEAR, described in Section 3. We generate a model for the latitudinally-varying H$_2$S and CH$_4$ abundance profiles that agree with the observations in each band in Section 4. In Section 5, we compare these results to other profiles of Neptune's trace gases and summarize how our findings impact our understanding of the dynamics and evolution of Neptune's upper atmosphere.

\section{Observations}
\label{S:2}

\subsection{Data}
\label{S:21}

We observed Neptune with ALMA, which is an interferometer located in the Atacama desert in northern Chile. A total of 66 high-precision antennas form the array: fifty-four 12m and twelve 7m. The arrangement of these antennas defines the angular resolution and maximum resolvable angular scale of the data. The tightest packed configuration, where the antennas are up to 150 m apart, allow large, faint objects to be observed. Extended arrangements, where the antennas are 16 km apart, provide a more detailed look. Our Neptune observations were taken between $2016-2017$ in Bands 3, 4, and 6, between $95-250$ GHz (1--3 mm). For each band, we selected the antenna configuration which allowed us to resolve Neptune and simultaneously see Neptune's entire (2.3'' diameter) disk. A typical resolution was 0.3'' using $\sim40$ 12-m antennas. We observed in four spectral windows in continuum mode covering Neptune's continuum spectra within each band. Table \ref{table1} summarizes the observations while Table \ref{table2} outlines the correlator and spectral setup. 

\begin{table}

\caption{Summary of observations.}
\begin{tabular}{llllll}
\hline
\\ \hline
 ALMA Band & Configuration & UT Date & Flux Calibrator & Bandpass Calibrator & T$_{\text{Source}}^{a}$ \\
 \hline
 Band 3 & C40-8 & 2017-07-27 & J0006-0623 & J2246-1206 &  726 \\
 Band 4 & C40-7 & 2016-10-07 & J2258-2758 & J2246-1206 & 1701 \\
 Band 6 & C40-7 & 2016-10-24 & J2258-2758 & J2246-1206 & 1032 \\
 \hline
\end{tabular}\label{table1}

\footnotesize{$^a$ Total science time on Neptune, in seconds} \\
\end{table}

\begin{table}

\caption{Summary of correlator setup, spectral windows, and synthesized beams.}

\begin{tabular}{lcccccc}
\hline
\\ \hline
ALMA Band & Center Frequency & Center Wavelength & Channel Width & Total Bandwidth$^a$ & Beam Size & Position Angle$^b$\\
& (GHz) & (mm) & (MHz) & (MHz) & (arcsec$^2$) & (degrees) \\ 
\hline
Band 3 & \begin{tabular}{@{}c@{}}95.012 \\ 96.970 \\ 107.000 \\ 109.000 \end{tabular} & \begin{tabular}{@{}c@{}}3.155 \\ 3.092 \\ 2.802 \\ 2.750 \end{tabular} & 15.625 & 2000 & \begin{tabular}{@{}c@{}}0.45"$\times$0.33" \\ 0.44"$\times$0.33" \\ 0.40"$\times$0.30" \\ 0.39"$\times$0.29"\end{tabular} & \begin{tabular}{@{}c@{}}$281.5^{\circ}$ \\ $282.1^{\circ}$\\ $282.9^{\circ}$ \\ $282.9^{\circ}$ \end{tabular} \\
\hline
Band 4 & \begin{tabular}{@{}c@{}}135.986\\ 137.924 \\ 147.986 \\ 149.986\end{tabular} & \begin{tabular}{@{}c@{}}2.205 \\ 2.174 \\ 2.026 \\ 1.999 \end{tabular} & 15.625 & 2000 & \begin{tabular}{@{}c@{}}0.38"$\times$0.28" \\ 0.38"$\times$0.28" \\ 0.35"$\times$0.26" \\ 0.35"$\times$0.26"\end{tabular} & \begin{tabular}{@{}c@{}}$23.0^{\circ}$ \\ $24.8^{\circ}$\\ $22.7^{\circ}$ \\ $24.3^{\circ}$ \end{tabular} \\
\hline
Band 6 & \begin{tabular}{@{}c@{}}223.982 \\ 225.982 \\ 239.981 \\ 241.981\end{tabular} & \begin{tabular}{@{}c@{}}1.338 \\ 1.327 \\ 1.249 \\ 1.239 \end{tabular} & 15.625 & 2000 & \begin{tabular}{@{}c@{}}0.31"$\times$0.25" \\ 0.32"$\times$0.24" \\ 0.30"$\times$0.23" \\ 0.29"$\times$0.23"\end{tabular} & \begin{tabular}{@{}c@{}}$64.3^{\circ}$ \\ $61.8^{\circ}$\\ $64.0^{\circ}$ \\ $63.6^{\circ}$ \end{tabular}\\
\hline
\end{tabular}\label{table2}

\footnotesize{$^a$ Each spectral window has the same total bandwidth, consisting of 128 channels with the same channel width.} \\
\footnotesize{$^b$ Defined as north through east.}

\end{table}

\subsection{Calibration and Imaging}
\label{S:22}
The obtained visibility data were reduced and calibrated in the Common Astronomical Software Application (CASA) version 5.1. We applied the standard ALMA pipeline to perform flagging (bad edge channels, shadowed antennas, and poor quality data), bandpass and flux calibration, and gain-time solutions. Table \ref{table1} lists our calibrator sources. Finally, we applied three iterations of self calibration on our Neptune data to remove short-term phase variability caused by fast atmospheric fluctuations \citep{Brogan2018}. The first iteration used the entire observation range while the second and third used 350-second and 60-second intervals. 

The multi-frequency synthesis mode in CASA \textit{tclean} was used to transform the visibility data into image maps. We used natural weighting and restricted clean to a circular mask that is roughly the size of Neptune's diameter plus twice the beam size. The natural weighting scheme gives equal weight to all baseline samples while the uniform weighting scheme gives equal weight to each spatial frequency. Since there are `naturally' more short baselines than long baselines, natural weighting preserves the peak sensitivity while uniform weighting reduces the peak sensitivity as the short baselines have been weighted less. In our testing, we found that uniform and intermediate weightings produced artificial speckles on the disk and large scale structures in the sky. Natural weighting limits these artifacts since the sensitivity is highest, but sacrifices some angular resolution since the long-baselines are under-weighted relative to the uniform scheme. Despite this, we still resolve roughly one-seventh of Neptune's disk. We also subtracted a limb-darkened model from the data to speed up the deconvolution process and reduce imaging artifacts. The number of clean iterations varied from 1000-2000, stopping once the noise within the planetary disk reached the noise in the sky.

Our resulting maps are 512$\times$512 pixels$^2$ with a cell size of 0.02". This cell size follows a common rule-of-thumb, which is to make the cell size roughly 1/10 the size of the synthesized beam. The planetary disk appears elliptically elongated due to convolution with the synthesized beam. This beam resembles a Gaussian with full-width at half maxima and position angles given in Table \ref{table2}.

\subsection{Error estimation}
\label{S:23}

 The error in our maps is calculated by averaging over four regions of the sky with boxes equal to the diameter of Neptune and taking the root-mean-square (RMS). RMS values range from $0.1-0.6$K. Table \ref{table3} lists our estimated errors in each band. This RMS does not include systematic effects, such as errors in the bandpass or flux calibration. The ALMA Calibrator Source Catalogue lists errors in the calibrator's flux estimate being about 5$\%$ or less in each band so we use this as an estimate for the absolute error in our disk-averaged temperature data (Section 5.1). 

\begin{table}

\caption{Summary of observed and modeled millimeter disk-averaged brightness temperatures.}
\begin{tabular}{llllll}
\hline
\\ \hline
Center Frequency (GHz) & Observed Flux Density$^a$ (Jy) & Observed T$_{\text{b}}^{a}$ (K) & Nominal T$_{\text{b}}^b$ (K) & Noise$^c$ (K) & Factor$^d$ \\
\hline
95.012 & $3.4\pm0.2$ & $126.6\pm6.3$ & 119.9 & 0.1 & 0.945 \\ 
96.970 & $3.5\pm0.2$ & $126.0\pm6.3$ & 119.4 & 0.1 & 0.950 \\ 
107.000 & $4.1\pm0.2$ & $120.5\pm6.0$ & 116.7 & 0.2 & 0.975 \\ 
109.000 & $4.1\pm0.2$& $118.8\pm6.0$ & 116.1 & 0.3 & 0.975 \\
\hline
135.986 & $5.9\pm0.3$ & $108.5\pm5.4$ & 109.0 & 0.3 & 1.010\\ 
137.924 & $6.1\pm0.3$ & $108.0\pm5.4$ & 108.5 & 0.2 & 1.010 \\  
147.986 & $6.7\pm0.3$ & $104.3\pm5.2$ & 105.7 & 0.2 & 1.020 \\ 
149.986 & $6.8\pm0.3$ & $104.5\pm5.2$ & 105.1 & 0.3 & 1.025 \\
\hline
223.982 & $13.3\pm0.7$ & $93.4\pm4.7$ & 94.8 & 0.4 & 1.020 \\ 
225.982 & $13.5\pm0.7$ & $93.0\pm4.7$ & 94.7 & 0.4 & 1.025 \\ 
239.981 & $15.1\pm0.8$ & $93.1\pm4.7$ & 93.4 & 0.6 & 1.010 \\ 
241.981 & $15.3\pm0.8$ & $92.8\pm4.6$ & 93.2 & 0.6 & 1.010 \\
\hline
\end{tabular}\label{table3}

\footnotesize{$^a$ The listed errors are the absolute errors, estimated at $5\%$ from the calibrators.} \\
\footnotesize{$^b$ The nominal model assumes a dry adiabat with trace gases enhanced by 30$\times$ the protosolar value.}\\
\footnotesize{$^c$ Random errors defined as the RMS on the sky.}\\
\footnotesize{$^d$ This factor is used to scale the observed brightness temperature to match the nominal model to within the estimated noise.}

\end{table}

\section{Models}
\label{S:4}

2D model maps of Neptune's disk were created using our radiative transfer (RT) code Radio-BEAR described in \citet{dePater2019}\footnote{This code is available at: \url{https://github.com/david-deboer/radiobear} .}. Given an atmospheric composition and thermal structure (described below), we calculate the RT-derived brightness temperatures of the planet on each location on the planet.

Radio-BEAR assumes that the atmosphere is in local thermodynamic equilibrium, where the temperature is calculated from deep in the atmosphere upwards assuming a dry or wet adiabat such that the temperature at 1 bar matches 71.5 K, that derived by \textit{Voyager} radio occultation measurements \citep{Lindal1992}. At altitudes above 1 bar, we use the temperature profile from \citet{Fletcher2010}. The temperature, pressure, and altitudes are related to each other through hydrostatic equilibrium. In all of our latitude-varying models (Sections 4.2 and 4.3), we assume that the temperature-pressure profile follows a dry adiabat in the troposphere. Cases of a wet adiabat are presented in the context of the disk-averaged brightness temperature in section 4.1. Our models also allow the abundance profiles of H$_2$S, CH$_4$ and \textit{ortho/para} H$_2$ to vary, as discussed below.

\begin{figure}
\centering
  \includegraphics[width=0.75\linewidth]{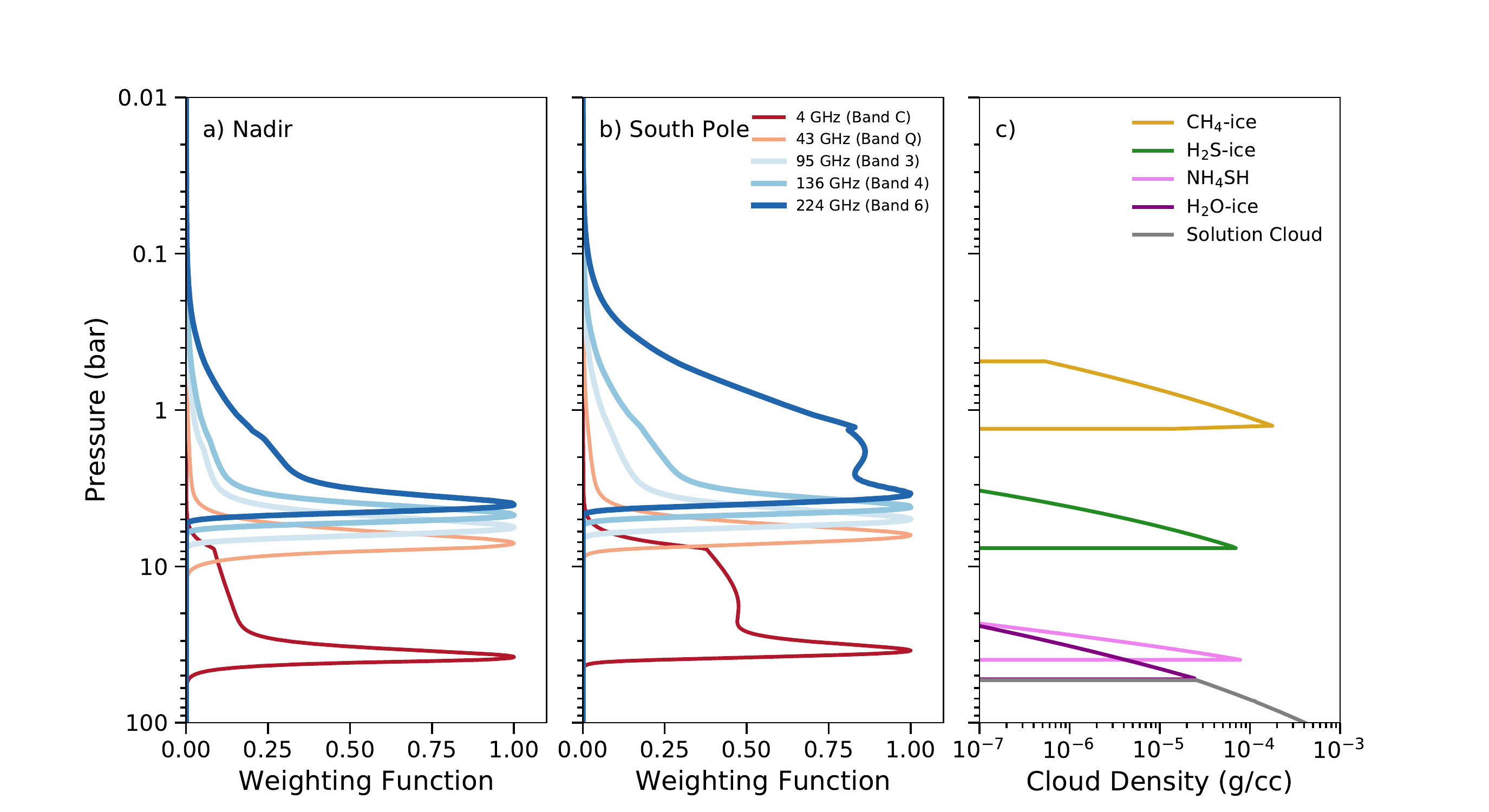}
  \caption{Normalized weighting functions at: a) nadir and b) the south pole, compared to the expected clouds expected to form on Neptune (c). The weighting functions at nadir and the south pole are both computed using the nominal abundance profile depicted in Figure \ref{fig:nomprofs}. Weighting functions are shown for representative frequencies in each ALMA band: 95 GHz (3.155 mm, Band 3), 136 GHz (2.205 mm, Band 4), and 224 GHz (1.338 mm, Band 6), as well as selected VLA frequencies: 4 and 43 GHz (6.2 cm and 0.7 cm, respectively). Note that the VLA frequencies probe significantly deeper into Neptune's atmosphere than the ALMA frequencies. The rightmost plot shows the density of different clouds expect to form on Neptune under thermochemical equilibrium.}
  \label{fig:contributions}
\end{figure}

Neptune's trace gases H$_2$S, CH$_4$, H$_2$O, and NH$_3$ in the deep atmosphere are assumed to be enhanced by $30\times$ their protosolar values\footnote{We use the protosolar values from \citet{Asplund2009}: C/H$_2$ = 5.90E-4; N/H$_2$ = 1.48E-4; O/H$_2$ = 1.07E-3; S/H$_2$ = 2.89E-5.} in our nominal model, apart from ammonia gas as 30$\times$ protosolar values for NH$_3$ are inconsistent with previous microwave disk-averaged temperatures \citep{dePater1985, dePater2014}. At higher altitudes, these gases follow the saturated vapor pressure curve with 100$\%$ relative humidity. Clouds expected to form under thermochemical equilibrium on Neptune include: an aqueous ammonia solution (H$_2$O-NH$_3$-H$_2$S) topped with water ice, ammonium-hydrosulfide (NH$_4$SH), and H$_2$S- and CH$_4$-ice \citep{Weidenschilling1973}. To form the NH$_4$SH cloud, H$_2$S and NH$_3$ are reduced in equal molar quantities until the product of their partial pressures reaches the equilibrium constant of the reaction forming NH$_4$SH. Once equilibrium is reached, only H$_2$S will remain to form clouds since there is practically no NH$_3$ gas remaining above the NH$_4$SH cloud. A tentative detection of H$_2$S spectral features near 1.58 $\mu$m in Neptune's troposphere implies that the deep bulk S/N ratio is greater than one \citep{Irwin2019}. The cloud density might affect microwave measurements \citep{dePater1991, dePater1993}. However, little is known about the cloud density on Neptune and the millimeter weighting functions peak at altitudes above the aqueous and ammonium-hydrosulfide clouds (see Fig. \ref{fig:contributions}). Clouds have also been shown to not affect the microwave opacity on Jupiter \citep{dePater2019}. Therefore, we ignore the effect of cloud opacity and focus on the gas opacity in our models. 

The gas opacity of Neptune's millimeter spectrum is dominated by H$_2$S and the collision-induced absorption (CIA) of H$_2$ (we include: H$_2$-H$_2$, H$_2$-He, H$_2$-CH$_4$). NH$_3$ and H$_2$O affect the spectra at wavelengths longer than 10 cm; NH$_3$ would have a larger impact at millimeter wavelengths if its abundance within Neptune were larger (see Fig. 4 of \citet{LuszczCook2013a}). The effect of PH$_3$ in the millimeter is most prominent at the ($1-0$) absorption line at 266.9 GHz. The width of this line is $\sim$ 20 GHz due to pressure broadening, meaning the wings of this absorption feature will have a small effect on the highest frequency data. However, this effect is well within the estimated noise of the maps so we do not add this gas in our models.

The \textit{ortho/para} H$_2$ fraction also influences Neptune's millimeter brightness temperature, by modifying both the adiabatic lapse rate and the gas opacity \citep{Trafton1967, Wallace1980, dePater1985, dePater1993}. The ratio of ortho to para hydrogen in equilibrium depends on temperature; however, fast vertical mixing could bring the ratio of ortho and para states of hydrogen away from equilibrium and towards a ``normal" ratio of 3 parts ortho to 1 part para. In this paper, we assume ``intermediate" H$_2$ proposed by \citet{Trafton1967} and used by e.g., \citet{LuszczCook2013b}: 
the ortho and para states of hydrogen (which define the CIA opacities) are set to the equilibrium value at the local temperature, while the specific heat is set to that of 'normal' hydrogen. For further explanation of intermediate hydrogen, see  e.g., \citet{Massie1982}. \citet{LuszczCook2013b} find that assuming a normal hydrogen fraction rather than equilibrium fraction significantly increases the opacity and lowers the brightness temperature by $5-6$K in their 1.2-mm model maps.

\begin{figure}
\centering
  \includegraphics[width=0.75\linewidth]{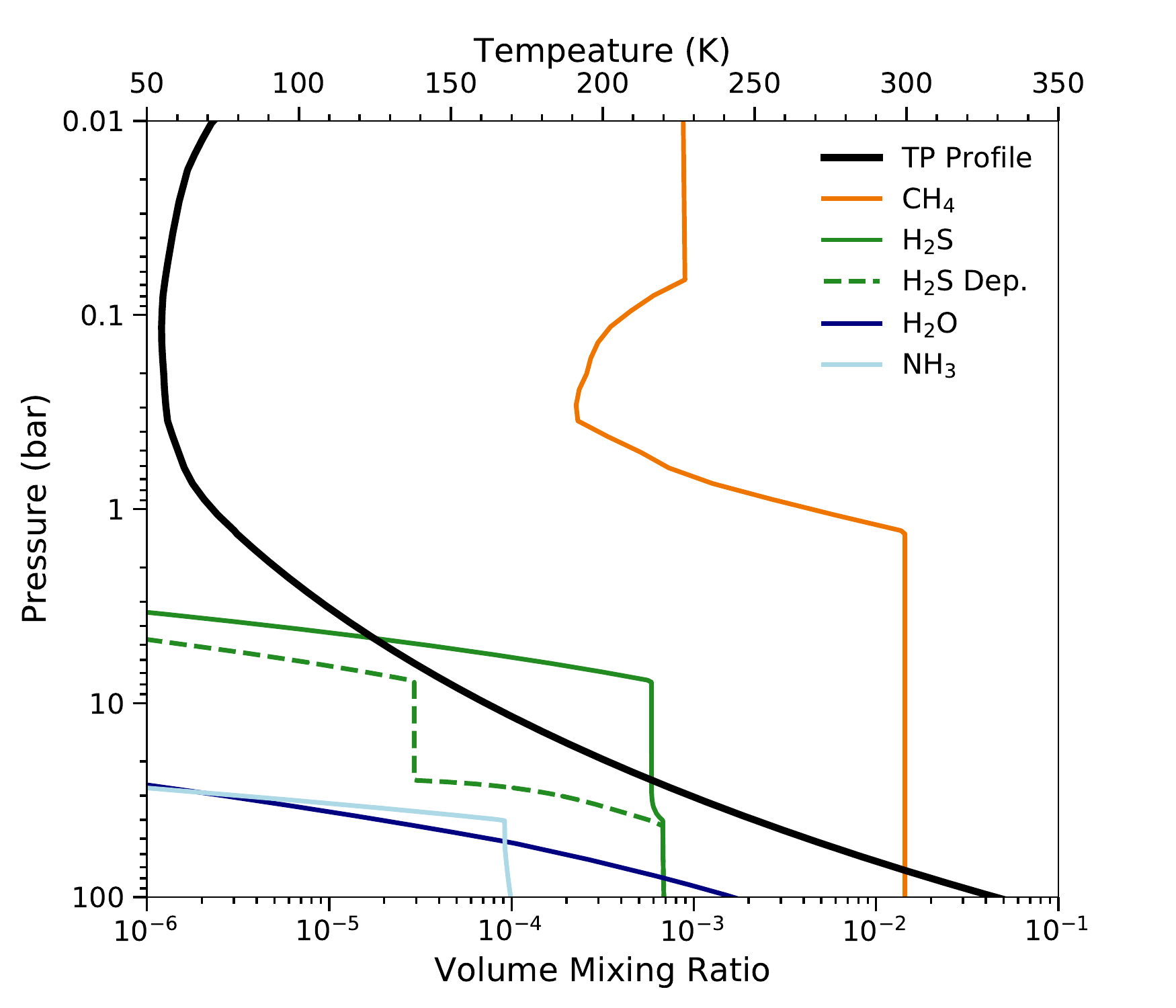}
  \caption{Temperature and abundance profiles of trace gases whose deep atmospheric abundances are enhanced by 30$\times$ their protosolar value, apart from NH$_3$ which is $1\times$ (solid lines). In the troposphere, the temperature follows a dry adiabat. These profiles define the nominal model. The dashed line is the H$_2$S depleted profile that that best fits Neptune's south pole from 2003 VLA data \citep{dePater2014}.}
  \label{fig:nomprofs}
\end{figure}

We generate 2D model maps of Neptune as follows: at the center frequencies of each spectral window, we calculate the RT-derived brightness temperatures of the planet on 9455 points on the disk and interpolate between these points to obtain the same resolution as our CASA imaged maps (0.02''). We then convolve each RT-map with an elliptical gaussian model of the synthesized beam. Our nominal model can be summarized as follows: 1) Neptune's trace gases are enhanced to 30$\times$ their protosolar value, except NH$_3$ which is $1\times$ protosolar; 2) `intermediate' \textit{ortho/para} H$_2$ (i.e., equilibrium \textit{ortho/para} H$_2$; specific heat close to that of normal H$_2$); 3) the temperature-pressure profile follows a dry adiabat in the troposphere. Figure \ref{fig:nomprofs} plots the temperature-pressure and abundance profiles for the nominal model. In the following section, we compare these beam-convolved model maps to the observed disk-averaged brightness temperatures and latitudinal brightness variations.  

\section{Results}
\label{S:5}

In the following subsections, we investigate the distribution of Neptune's observed brightness temperature. First, we look at the disk-averaged brightness temperature and compute a variety of RT models using dry and wet adiabats and different enhancements of the deep abundance of trace gases. From these results, we present our nominal model (the same as that described at the end of Section 3), which fits both our ALMA and 2003 VLA disk-averaged temperatures simultaneously. Second, we subtract our nominal model from the data to produce residual maps. These maps show seven distinct latitudinal bands across Neptune's disk. In the final subsections, we find abundance profiles of H$_2$S, CH$_4$, and \textit{ortho/para} H$_2$ which fit the observed brightness variations in each latitudinal band. As an example of how we do this, we first look at Neptune's south polar cap, where previous VLA studies constrained the deep abundance of H$_2$S. This provides a good litmus test for our model and fitting routine. We then move on to the rest of Neptune's disk and show that we can fit the observed brightness temperature distribution at all latitude bands by varying the abundances of the aforementioned species. 

\subsection{Disk-Averaged Brightness Temperatures}
\label{S:5.1}

We calculate Neptune's disk-averaged brightness temperature from the ALMA data in two ways. First, we sum the flux density contained within the planetary disk convolved with the model beam. Second, we fit the u-v short-spacing amplitude with a Bessel function and obtain an estimate of the zero-spacing flux density. Both results agree to well within the absolute calibration errors. We report the average of these two methods in Table \ref{table3}. Since Neptune blocks the cosmic microwave background (CMB), its true brightness temperature is higher than observed in the radio. In all of our millimeter observations, we correct for the CMB by following the procedure laid out in Appendix A of \citet{dePater2014}.

Figure \ref{fig:diskavgtemp} plots our observed radio disk-averaged brightness temperatures on top of RT model spectra. In this plot, we show the effect of varying the deep abundance enhancement of trace gases (H$_2$S, CH$_4$, H$_2$O, and NH$_3$) relative to their protosolar values and consider both wet and dry lapse rates. We find that the nominal model: 30$\times$ enhancement with a dry lapse rate, agrees very well with both the ALMA millimeter and the 2003 VLA centimeter data from \citet{dePater2014}. Models with temperatures following wet lapse rates are too cold relative to these data. We acknowledge, though, the following shortcomings in our models. First, these models are highly degenerate. Gas abundances are seldom at 100$\%$ humidity and so a profile with a wet adiabat where the gases are sub-saturated may result in high enough brightness temperatures to match the data. Second, our models do not account for the cooling (heating) of the atmosphere from adiabatic expansion (compression) at latitudes where the air is rising (sinking). Finally, apart from the dry and wet adiabat, we do not consider latitudinal variations in temperature. \citet{Conrath1998} and \citet{Orton2007} see $2-3$ K latitudinal temperature variations between $50-100$ mbar. These temperature variations are similar in strength to those we see in our residual maps (Fig. \ref{fig:residualmaps}), particularly in Band 6. \citet{LuszczCook2013b} created temperature profiles matching the $2-3$ K variations described in \citet{Conrath1998} and \citet{Orton2007} at altitudes above 1 bar. These temperature-varying models reduced the overall $\chi^2$ in their fit to 1.2 mm continuum data. However, we expect similar profiles to have only a modest effect in Band 6, where the contribution functions peak at the highest altitudes (but still below 1 bar, see Fig. \ref{fig:contributions}), and little effect in Bands 3 and 4, which probe pressures much higher than where temperature variations are seen in the infrared. Below 1 bar, the temperature profile should strictly follow an adiabat, regardless of latitude. In summary, while we do not consider temperature variations in our analysis, we do not expect them to be the primary cause of the observed brightness temperature distribution.

\begin{figure}
\centering
  \includegraphics[width=0.75\linewidth]{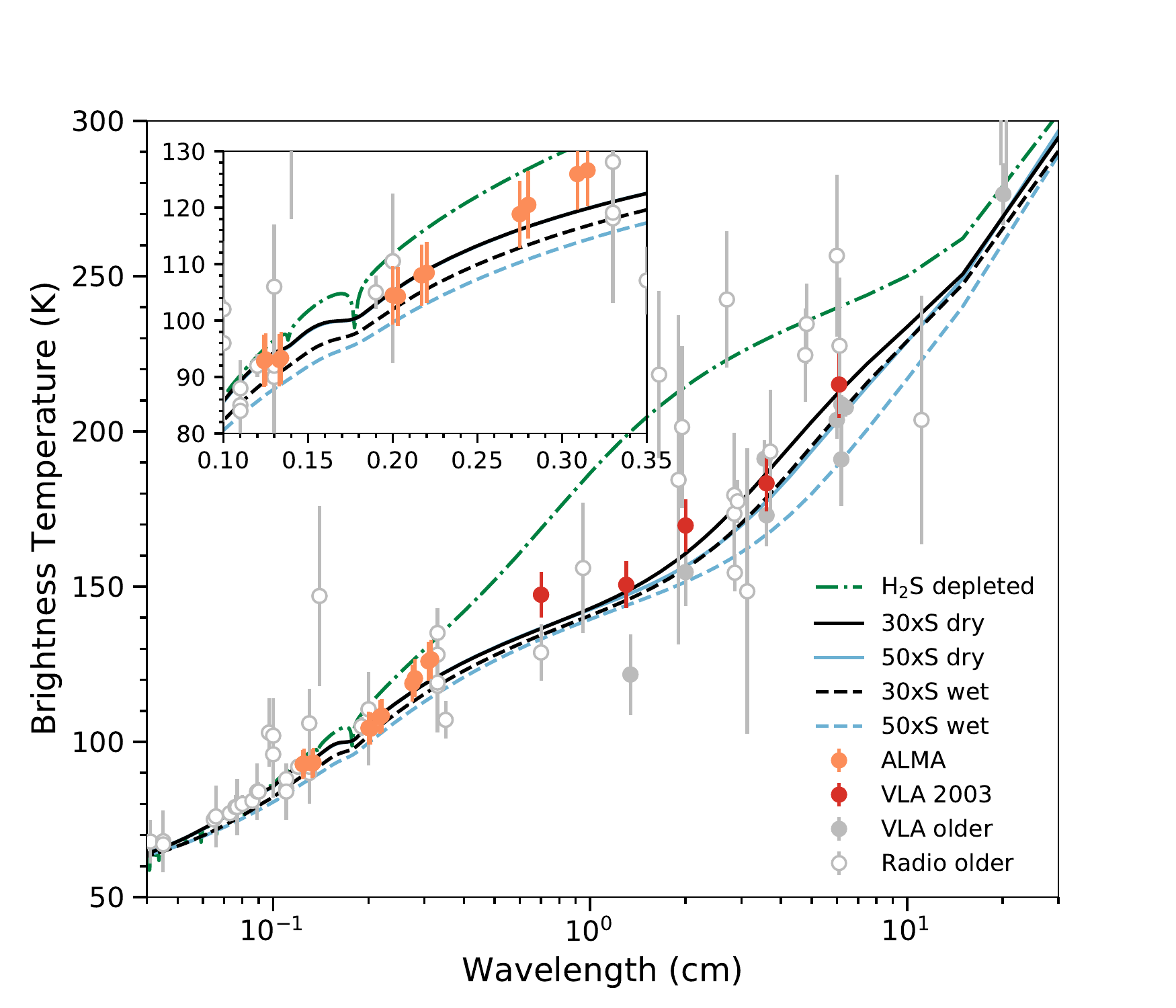}
  \caption{Disk-averaged brightness temperatures of Neptune. The ALMA data are plotted as orange points, with 5\% absolute errors estimated from the calibrators. In addition, VLA 2003 data are plotted in red \citep{dePater2014}, along with older VLA data in gray \citep{dePater1991} and older single dish radio data in open circles \citep{dePater1989}. Overplotted are model spectra which vary the deep abundance of H$_2$S, CH$_4$, and H$_2$O (30$\times$ and 50$\times$ their protosolar values in black and light blue respectively), and compare dry and wet lapse rates (solid and dashed lines, respectively). A model for Neptune's south polar hot spot which depletes H$_2$S above 43 bar is also shown in the dot-dash green line \citep{dePater2014}.}
  \label{fig:diskavgtemp}
\end{figure}

\subsection{Latitudinal Brightness Temperature Variations}
\label{S:5.2}

\begin{figure}
\centering
  \includegraphics[width=0.75\linewidth]{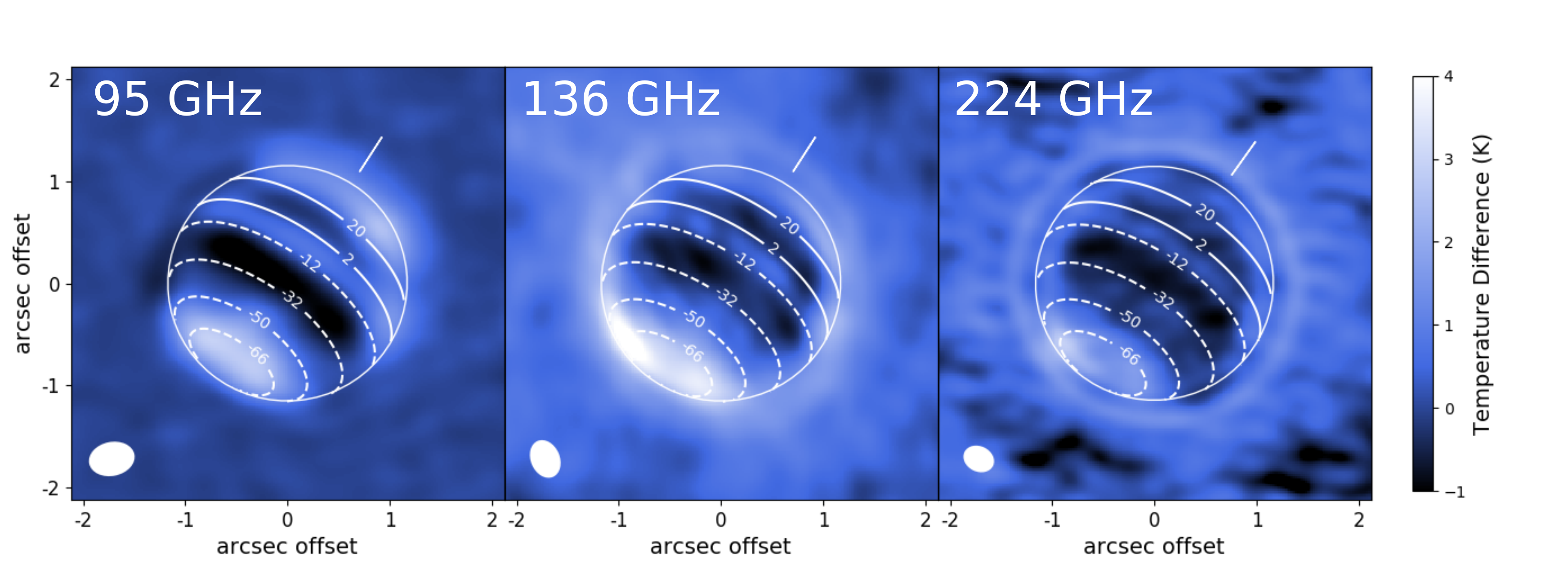}
  \caption{ALMA residual maps where the beam-convolved nominal model has been subtracted from the data. Contour lines delineate the latitude transitions between bands. Dark bands represent cold brightness temperatures relative to the model, while bright bands are warmer than the model. The FWHM of the beam is indicated in white in the bottom left of each map.}
  \label{fig:residualmaps}
\end{figure}

In Figure \ref{fig:residualmaps}, we show residual maps of Neptune in the first spectral window of each ALMA band. Since the absolute calibration is imperfect, each map is scaled by a factor such that the observed disk-averaged temperature matches the nominal model (see Table 3). As a result, our main assumption in this analysis is that Neptune's disk-average matches the nominal model. Deviations from this model result in the latitudinal structure evident in the residuals. This structure is due to changes in the brightness temperature that we assume to be due to variations in the opacity. The dark areas at the southern mid-latitudes and in the northern equatorial region are interpreted as probing higher, colder altitudes due to enhancements in absorbers. Conversely, the south pole appears bright in the radio due to opacity depletions, allowing the deeper, warmer layers to be probed. 

\begin{figure}
\centering
  \includegraphics[width=0.75\linewidth]{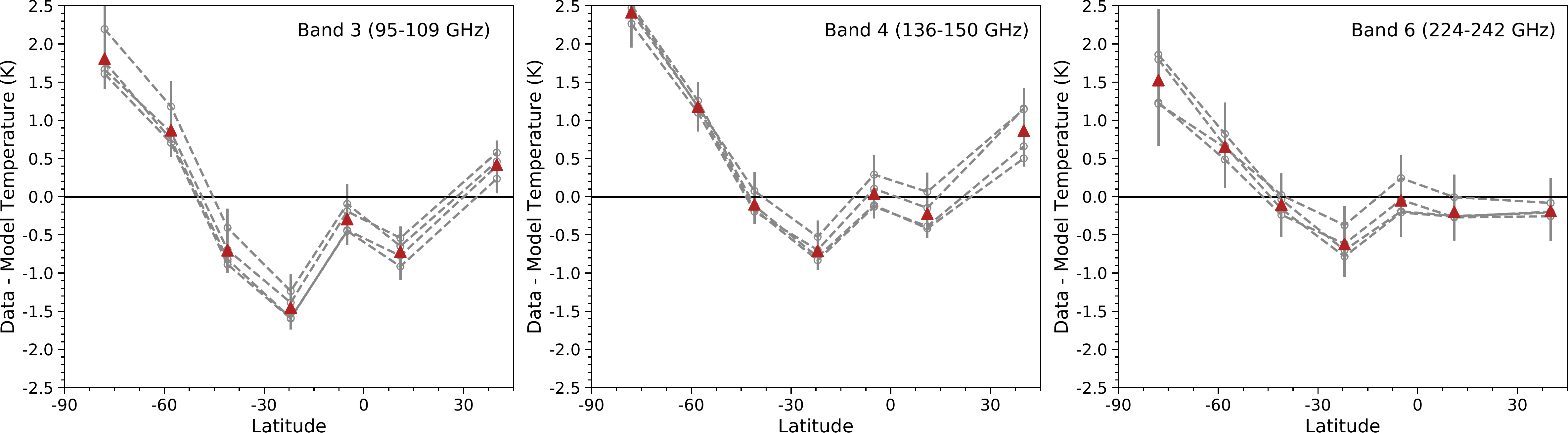}
  \caption{Residual temperatures comparing the data to the beam-convolved nominal model versus latitude (Band 3 left, Band 4 middle, Band 6 right). Points are plotted at the central latitude in the seven identified bins. The four gray dashed lines are the residual temperatures for each of the four spectral windows in a band. Red triangles are the average residual over each spectral window. The residuals in each latitude bin are calculated by averaging over each pixel within 60$^{\circ}$ of the sub-observer longitude. Error bars are the image noise divided by the square-root of the number of ALMA beams that fit into the corresponding bin.}
  \label{fig:residuals_h2s30}
\end{figure}

Planetary coordinates are computed for each pixel on the disk using ephemeris data from JPL Horizons, with the center pixel equaling the sub-observer latitude and longitude. Latitudes are reported in planetographic coordinates. We identify seven bands on Neptune that correspond to discrete changes in the temperature structure: 90$^{\circ}$S$-66^{\circ}$S, 66$^{\circ}$S$-55^{\circ}$S, 55$^{\circ}$S$-32^{\circ}$S, 32$^{\circ}$S$-12^{\circ}$S, 12$^{\circ}$S$-2^{\circ}$N, 2$^{\circ}$N$-20^{\circ}$N, northward of 20$^{\circ}$N. Each latitude bin is well-resolved, covering at least the total area of the ALMA beam, and is in planetographic coordinates. In Figure \ref{fig:residuals_h2s30}, we plot the average temperature difference between the data and nominal model per spectral window and band. Differences that are twice the RMS noise reported in Table 3 are significant.

\subsection{Neptune's South Polar Cap}
\label{S:5.4}

\begin{figure}
\centering
  \includegraphics[width=0.75\linewidth]{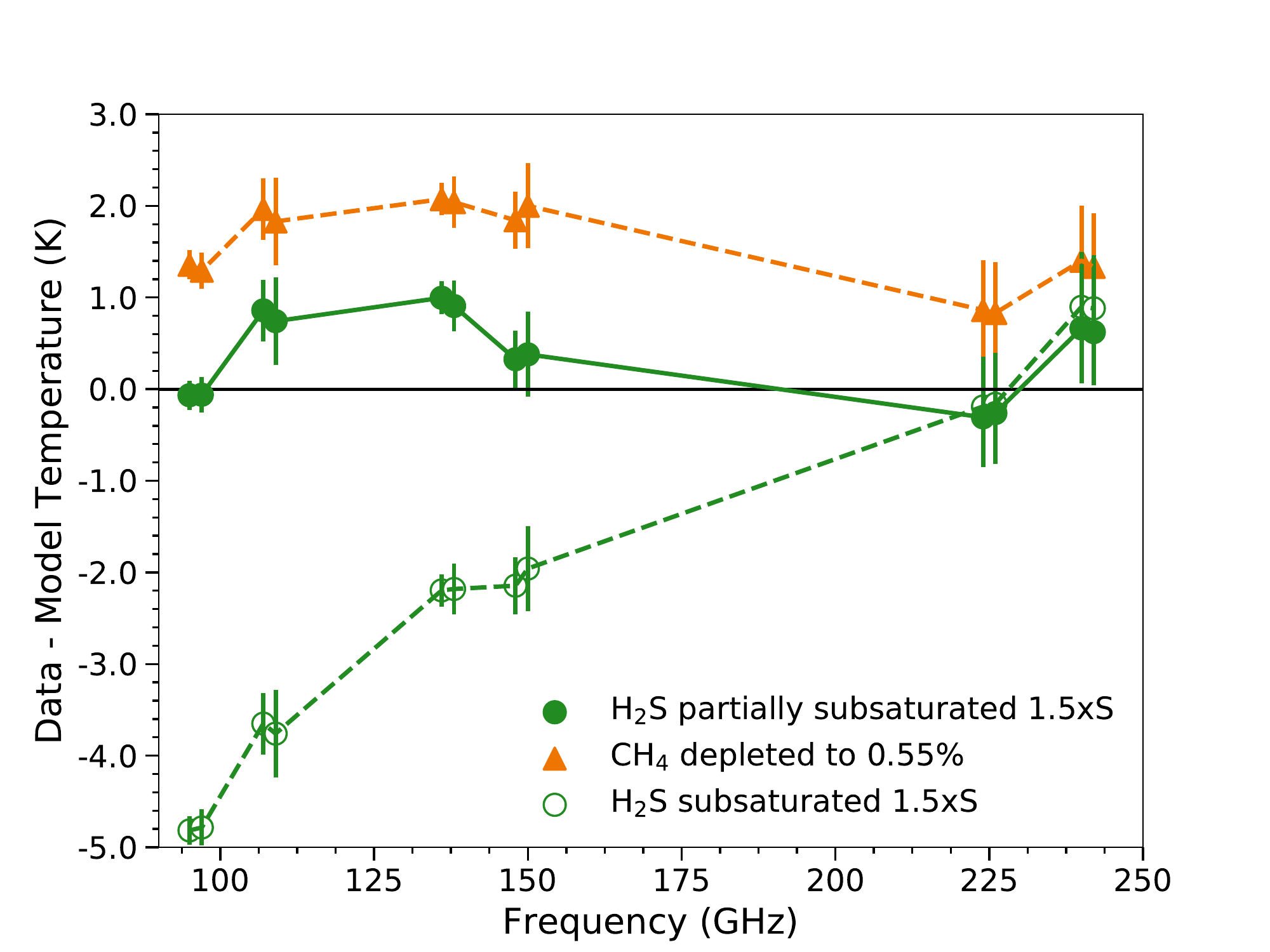}
  \caption{Residuals of the south polar data minus each of three different models, as described in Section 4.3. The dashed green line represents the residuals for the data compared with the model assuming H$_2$S subsaturation to $5\%$ of its nominal value, as suggested by the VLA cm data in \citet{dePater2014}. The orange dashed line and orange triangles are residuals to the south pole region in a model which only depletes CH$_4$ to 0.55$\%$, holding the other trace gases to their nominal values. The solid green line represents the residuals for the data compared with a model in which H$_2$S is partially subsaturated. This model depletes H$_2$S to 1.5$\times$S above 43 bar before following the saturation vapor pressure curve up to 4.5 bar and becoming subsaturated at higher altitudes (see Figure \ref{fig:sp-profiles} for a plot of this profile). Simultaneously, CH$_4$ is depleted to 0.55$\%$ in the deep atmosphere. This is our preferred model and is consistent with the observations of \citet{dePater2014} and \citet{LuszczCook2013b}. }
  \label{fig:sp-residuals}
\end{figure}

The ALMA residual maps show clear warming on Neptune at latitudes southward of -66$^{\circ}$. This region is Neptune's south polar cap and has been detected as a hot spot in the radio \citep{Butler2012DPS, LuszczCook2013b, dePater2014} and in the mid-infrared \citep{Hammel2007b, Orton2007}. \citet{dePater2014} found a best fit to their VLA data to the hot spot by defining a `plateau' of constant low opacity from 66$^{\circ}$S to the south pole, depleting H$_2$S to 1.5$\times$ the protosolar value (or 5$\%$ their nominal 50$\times$ model) above 43 bar (i.e., above the NH$_4$SH cloud). The deep CH$_4$ abundance is also depleted to 1.5$\times$ the protosolar value in their model (equivalently, a 0.072$\%$ volume mixing ratio), but itself is not a source of opacity at cm-wavelengths and so this study did not attempt to constrain the south pole CH$_4$ content (although methane does affect the adiabat and opacity due to CIA). \citet{LuszczCook2013b} found that the high southern latitudes in their CARMA 1.3-mm map were consistent with the VLA model. In Figure \ref{fig:nomprofs}, we show this depleted profile and in Figure \ref{fig:sp-residuals} we compare the bin-averaged residual brightness temperatures for the ALMA observations and the de Pater VLA depleted model. We find good agreement in Band 6, which covers the frequency of the CARMA map, however the model is too bright at lower frequencies. \citet{LuszczCook2013b} found that the high southern latitudes in their CARMA data could be matched if CH$_4$ were depleted to 0.55$\%$, while keeping H$_2$S at the nominal value. A comparison of our data with this CH$_4$ depleted model is also shown in Figure \ref{fig:sp-residuals}. Once again, the Band 6 data agree with the 1.3-mm map of Neptune presented in \citet{LuszczCook2013b} to within their estimated uncertainties. However, the model is too cold to match the new low frequency data. 

\begin{figure}
\centering
  \includegraphics[width=0.75\linewidth]{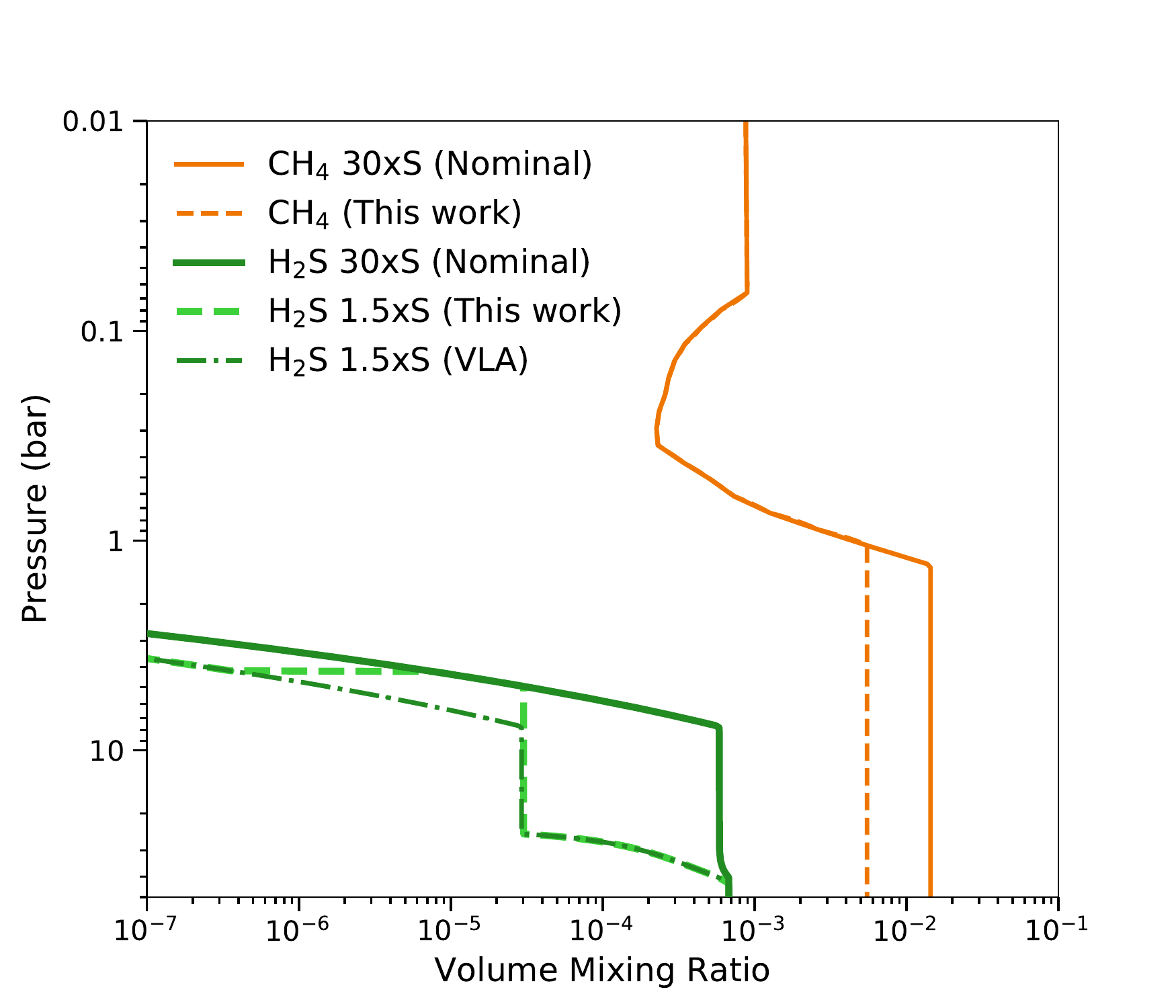}
  \caption{H$_2$S (green) and CH$_4$ (orange) abundance profiles. Models depleting one or both of these gases are used to explain radio brightness enhancements at Neptune's South Pole. The 30$\times$S nominal profiles are shown as solid lines. The H$_2$S model fitting VLA measurements depletes H$_2$S down to $5\%$ (1.5$\times$S) of the nominal profile at $P < 43$ bar; this profile is plotted as a dot-dash green line \citep{dePater2014}. Our models for the H$_2$S and CH$_4$ profile, which simultaneously fits the ALMA, CARMA, and VLA measurements, are plotted as dashed lines. Note that in our H$_2S$ profile, H$_2$S partially follows the saturation pressure curve, and follows the subsaturated profile of \citealp{dePater2014} higher than 4.5 bar.}
  \label{fig:sp-profiles}
\end{figure}

A better fit at low frequencies is obtained by setting the deep atmospheric CH$_4$ abundance to 0.55$\%$ and by depleting H$_2$S to 1.5$\times$ protosolar from 43 bar up to the saturated vapor pressure curve. The H$_2$S abundance follows the saturated vapor pressure curve below 4.5 bar; for $P < 4.5$ bar, H$_2$S is subsaturated down to $5\%$ of the saturated vapor curve. This adjusted profile is plotted in Figure \ref{fig:sp-profiles} and the temperature residuals are shown in Figure \ref{fig:sp-residuals}. We emphasize that some subsaturation is needed in order to fit the data well. Since the millimeter is most sensitive to the pressures where H$_2$S saturation occurs, all models in which H$_2$S profiles follow the saturated vapor pressure curve appear indistinguishable in terms of their contributions to the overall opacity and all under-predict the brightness temperatures in this region. Therefore, even models significantly depleting the deep abundance of H$_2$S will have temperatures that are too cold at the south pole and will resemble the residuals in Figure \ref{fig:residuals_h2s30}.

With this new depleted model for Neptune's south polar hot spot, we can fit the VLA, CARMA, and ALMA data simultaneously. This is because the VLA cm-data probe pressures deeper than our alterations at the H$_2$S saturation curve and we have not changed the deep H$_2$S abundance estimated from \citet{dePater2014} nor the deep CH$_4$ abundance determined by \citet{LuszczCook2013b}. The CARMA maps have larger uncertainties ($2-4$ K) than our maps presented here, and so depleting both H$_2$S and CH$_4$ is consistent with those data.

\subsection{Constraining Neptune's Variations with Latitude}
\label{S:5.3}

\begin{figure}
\centering
  \includegraphics[width=0.75\linewidth]{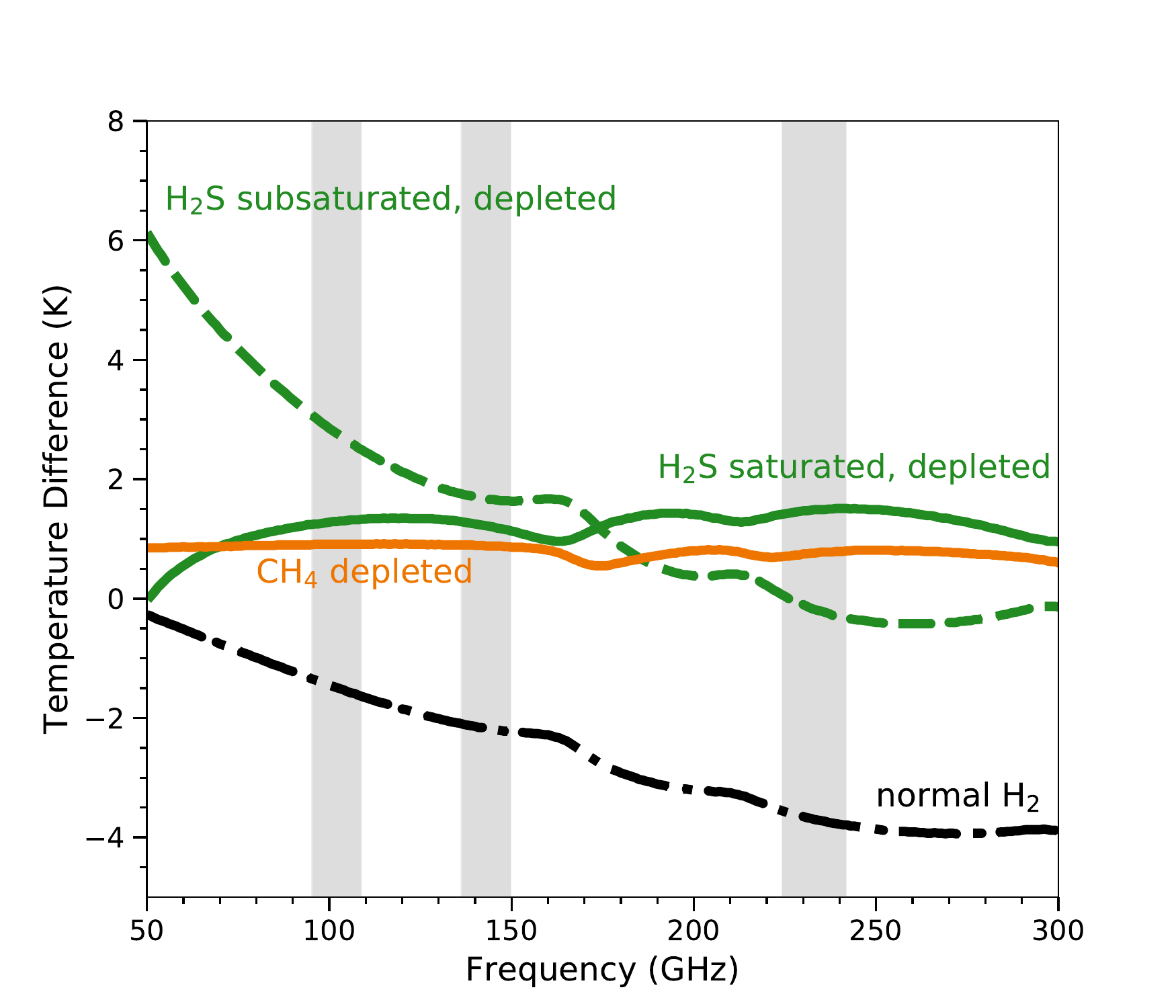}
  \caption{A comparison of models of Neptune's spectrum to gauge how different opacity sources affect the brightness temperature across millimeter wavelengths. Plotted is the disk-averaged  brightness temperatures for various models with the nominal model subtracted. The orange model depletes CH$_4$ to 15$\times$ the protosolar value (0.72$\%$) below saturation and the follows saturated vapor pressure curve above (as in the dashed-orange line in Fig. \ref{fig:sp-profiles}). The solid green model depletes H$_2$S to 15$\times$ the protosolar value (3.52E-4 mixing ratio) in the deep atmosphere, but follows the saturation curve. The dashed green model depletes H$_2$S similarly, but subsaturates H$_2$S by the same fraction (similar to the dashed green profile in Figure \ref{fig:nomprofs}). The black dot-dash model substitutes opacity by normal hydrogen for equilibrium hydrogen. Gray rectangles indicate the ALMA bands (Bands 3, 4, and 6 left-to-right).}
  \label{fig:comp_vs_spectra}
\end{figure}

From Figures \ref{fig:residualmaps} and \ref{fig:residuals_h2s30}, it is clear that the magnitude of temperature variations across Neptune's disk varies with wavelength. In the millimeter, these variations are caused by altering the abundance of trace gases: H$_2$S and CH$_4$, or changing the fraction of \textit{ortho/para} H$_2$. Figure \ref{fig:comp_vs_spectra} compares the millimeter-spectrum of the nominal model to that of models depleted in trace gases. These plots show that the spectral shape varies with both wavelength and composition. Models where the H$_2$S profile follows the saturated vapor pressure curve differ little from the nominal model, no matter the abundance alterations in the deep troposphere. This is because the ALMA bands probe altitudes above 10 bar, where H$_2$S saturation begins, meaning changes in the constant deep abundance profile are undetected. On the other hand, models which subsaturate H$_2$S significantly increase the brightness temperature at short frequencies. The Band 6 (high frequencies) contribution functions peak between $1-5$ bar while Band 3 (low frequencies) probes between $5-10$ bar, altitudes close to where H$_2$S begins saturation. Thus, subsaturating H$_2$S results in a larger loss of opacity at low frequencies than the frequencies which probe higher altitudes, meaning low frequencies are able to probe deeper, warmer layers. In contrast, depleting CH$_4$ results in uniform increases across the millimeter. Substituting normal hydrogen for equilibrium hydrogen decreases the brightness temperature by 4K in Band 6 and less than 2K in Band 3. These differences with wavelength enable us to disentangle the effect of each constituent on the observed brightness temperatures in each spectral band. 

In the following subsections, we present models where we vary a single constituent: H$_2$S, CH$_4$, or \textit{ortho/para} H$_2$, while holding the others at their nominal value. We find the profiles which best match the data in each latitude bin for a single band. Then, we compare how well this matches the data in the other spectral bands. This gives a sense of how important the varied constituent is in producing model temperatures for each observation. Following this procedure, we present our best model fits to all three spectral bands where every parameter is allowed to vary. Table \ref{table:modeldesc} describes how the constituent profiles change over all latitude and models.

The significance of our results within a particular latitude bin is computed with the reduced $\chi^2$:
\begin{equation}
    \chi^2 = \frac{1}{M-N}\sum_{m=0}^{M-1} \frac{\delta T_m^2}{\sigma_m^2},
\end{equation}
where $\delta T_m$ is the difference between the data and model brightness temperature at each spectral window $m$, $\sigma_m$ is the image noise (see Table 3) divided by the square root of the number of beams that fit within the particular latitude bin. $M - N$ is the number of degrees of freedom in the model. We calculate the probability $p$ that this value of the reduced $\chi^2$, or a larger one, could arise by chance given $M-N$. We take the number of degrees of freedom to be eight: twelve spectral windows ($M$) minus four free parameters ($N$): the deep abundances of H$_2$S and CH$_4$, the \textit{ortho/para} H$_2$ fraction, and the scaling factor assumption. For eight degrees of freedom, $p < 0.05$ when the reduced $\chi^2 > 15.5$. Therefore, models with reduced $\chi^2 > 15.5$ are unlikely to occur due to random chance and are inconsistent with the observations. Table \ref{table:chisq} summarizes the various fits to the data across each latitude range.

\subsubsection{Varying H$_2$S}
\begin{figure}
\centering
  \includegraphics[width=0.75\linewidth]{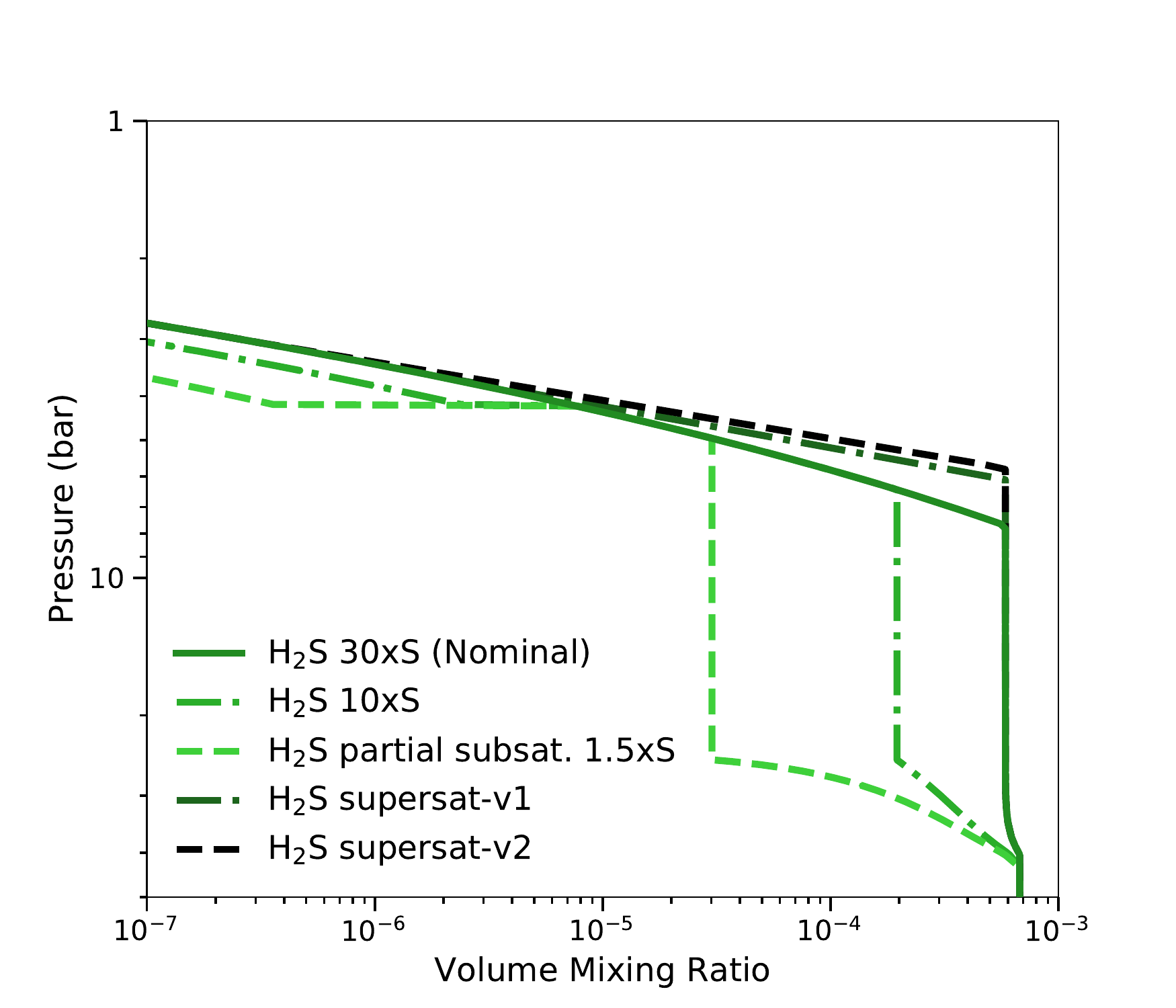}
  \caption{Examples of H$_2$S abundance profiles used to best fit the ALMA data. The light green dashed line and solid green line are the same profiles as depicted in Fig. \ref{fig:sp-profiles}. The dark green and black dashed and dot-dashed lines are supersaturated profiles (abbreviated supersat-v1 and supersat-v2).}
  \label{fig:h2sprof}
\end{figure}

We first consider variations in H$_2$S that best match the Band 3 latitude variations. For each latitude band other than the south pole, we produce a grid of models in which the H$_2$S mixing ratio above 43 bar (the NH$_4$SH cloud) is set from $5-30\times$ solar in 5$\times$S steps. The profile then follows the saturation curve up to 4.5 bar, and is either depleted above 4.5 bar or supersaturated to a higher altitude. We choose 4.5 bar as this transition pressure as this matches the data at the south pole best. For supersaturated models, the H$_2$S abundance is set to the constant deep atmospheric value until $5-7$ bar, testing a grid of models in 0.25 bar steps. At altitudes above 3 bar (the high-altitude limit of H$_2$S cloud formation), the H$_2$S profile follows the saturation curve. At pressures in between 3 bar and $5-7$ bar, the abundance is assumed to be linear in log-log space. For the south pole, we do not produce new model fits but continue to use the model from Section 4.3, which is depleted to a much higher depth to be consistent with VLA observations. The deep CH$_4$ abundance is held to its nominal value of 1.44$\%$ ($30\times$ S) and equilibrium H$_2$ is assumed. Figure \ref{fig:h2sprof} plots example H$_2$S profiles used in these fits.

\begin{figure}
\centering
  \includegraphics[width=0.75\linewidth]{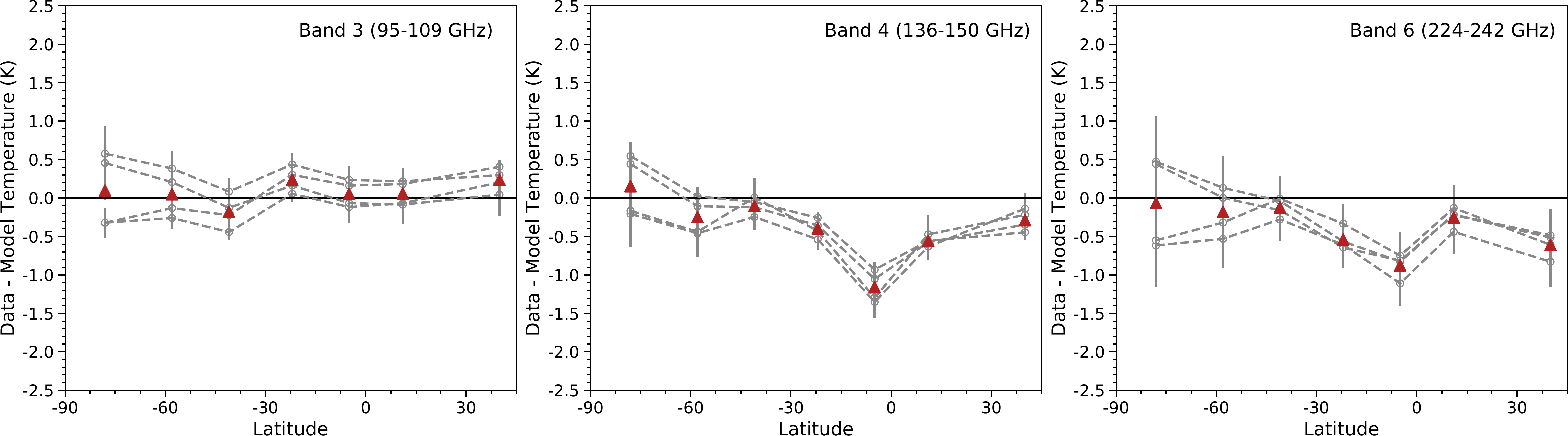}
  \caption{Residual temperatures comparing the data to the beam-convolved nominal model versus latitude, as in Fig. \ref{fig:residuals_h2s30}, allowing H$_2$S (only) to vary with latitude to best-match the Band 3 data.}
  \label{fig:residuals_h2s}
\end{figure}

From the residuals plotted in Figure \ref{fig:residuals_h2s}, there is general good agreement between the data and model in each band, apart from 12$^{\circ}$S$-2^{\circ}$N where the model in Bands 4 and 6 is too warm compared to the data. This suggests the need to enhance CH$_4$ or add normal H$_2$ in this area.

\subsubsection{Varying CH$_4$}
In order to determine if variations in CH$_4$ may effect the brightness temperature, we find CH$_4$ profiles which fit Band 6 to within the error bars of the average residual of the four spectral windows. The H$_2$S abundance is held at the nominal 30$\times$S profile and equilibrium H$_2$ is assumed. CH$_4$ is either enhanced or depleted from the nominal volume mixing ratio (1.44$\%$ or 30$\times$S), below the saturation curve. We create a grid of models in mixing ratio step-sizes of 0.36$\%$ (or 7.5$\times$S), testing deep abundances between $0.55-4.4\%$, the limits considered in \citealp{LuszczCook2013b}.

\begin{figure}
\centering
  \includegraphics[width=0.75\linewidth]{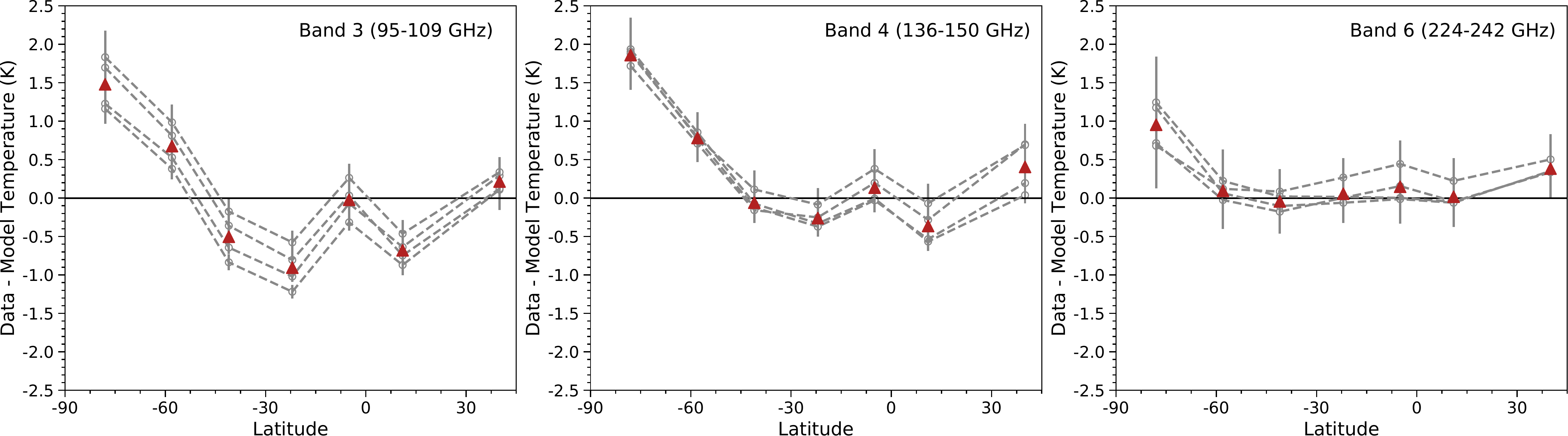}
  \caption{As Fig. \ref{fig:residuals_h2s}, but allowing only CH$_4$ to vary with latitude. We find CH$_4$ profiles that match the Band 6 data and compare these with the Band 3 and Band 4 data.}
  \label{fig:residuals_ch4}
\end{figure}

The residuals in each band are plotted in Figure \ref{fig:residuals_ch4}. Bands 3 and 4 do not fit a CH$_4$-varying only model well, implying H$_2$S must vary as well. 

\subsubsection{Varying \textit{ortho/para} H$_2$}
\begin{figure}
\centering
  \includegraphics[width=0.75\linewidth]{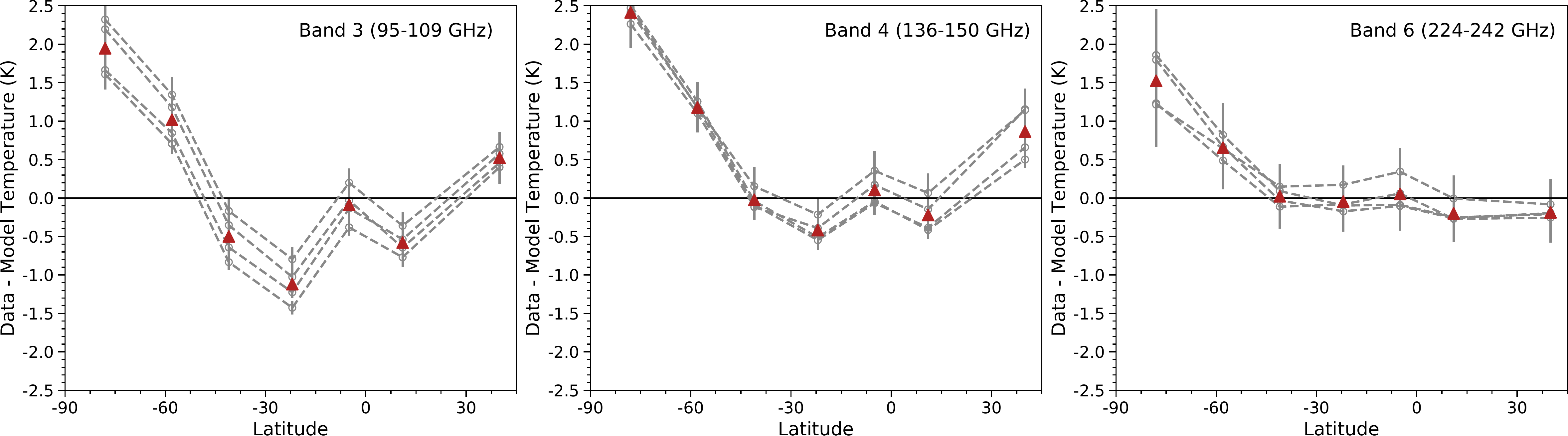}
  \caption{As Fig. \ref{fig:residuals_h2s}, but allowing only the fraction of equilibrium H$_2$ to vary with latitude. We find H$_2$ profiles that match the Band 6 data and compare these with the Band 3 and Band 4 data.}
  \label{fig:residuals_h2n}
\end{figure}

In these fits, we consider the fraction of equilibrium H$_2$ to normal H$_2$ needed to match the Band 6 brightness temperatures. Since normal H$_2$ decreases the brightness temperature, adding normal H$_2$ to latitudes with positive temperature residuals (like the south pole) is implausible. Figure \ref{fig:residuals_h2n} plots the temperature residuals in each band. We find that a hydrogen mixture of 90$\%$ equilibrium hydrogen and 10$\%$ normal hydrogen can explain the Band 6 data between 32$^{\circ}$S$-12^{\circ}$S, with equilibrium H$_2$ used everywhere else. However, the other ALMA spectral bands fail to fit the data from 32$^{\circ}$S$-12^{\circ}$S. This result is expected based on the spectral analysis in figure \ref{fig:comp_vs_spectra}, showing the minimal effect normal hydrogen has in Band 3. For the best fit model, we therefore do not consider normal hydrogen. We emphasize however that our results do not preclude a small fraction of normal H$_2$, on the order of $\leq 10\%$, existing at the pressures probed by ALMA. This is consistent with findings from \textit{Voyager} IRIS measurements (\citealp{Conrath1998, Fletcher2014}. These studies find that the para H$_2$ fraction deviates from its expected equilibrium values on the order of $2-5\%$ between $0.01-1$ bar. 

\subsubsection{Best Fit Model}
\begin{figure}%
\centering
\subfigure{\includegraphics[width=0.4\linewidth]{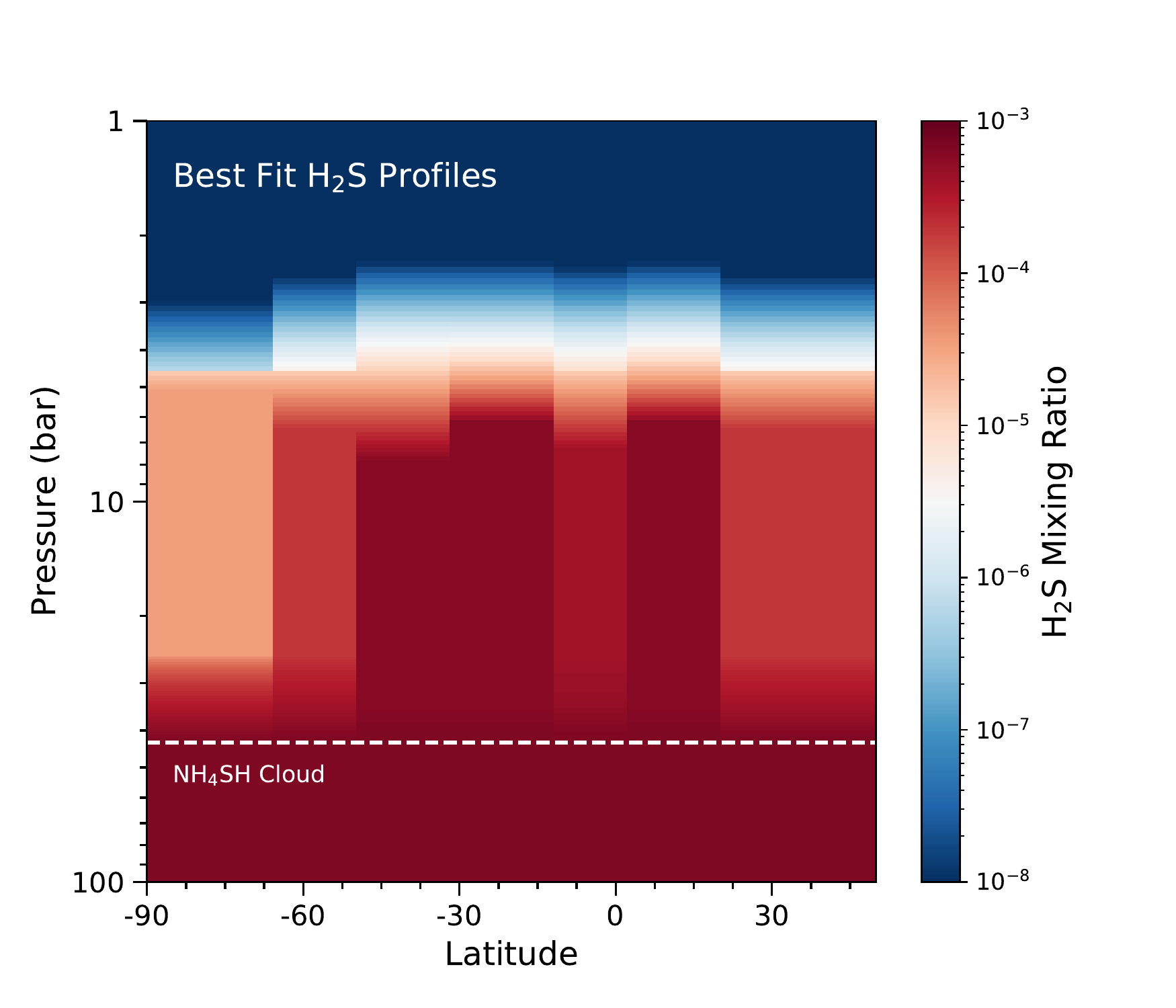}}%
\subfigure{\includegraphics[width=0.4\linewidth]{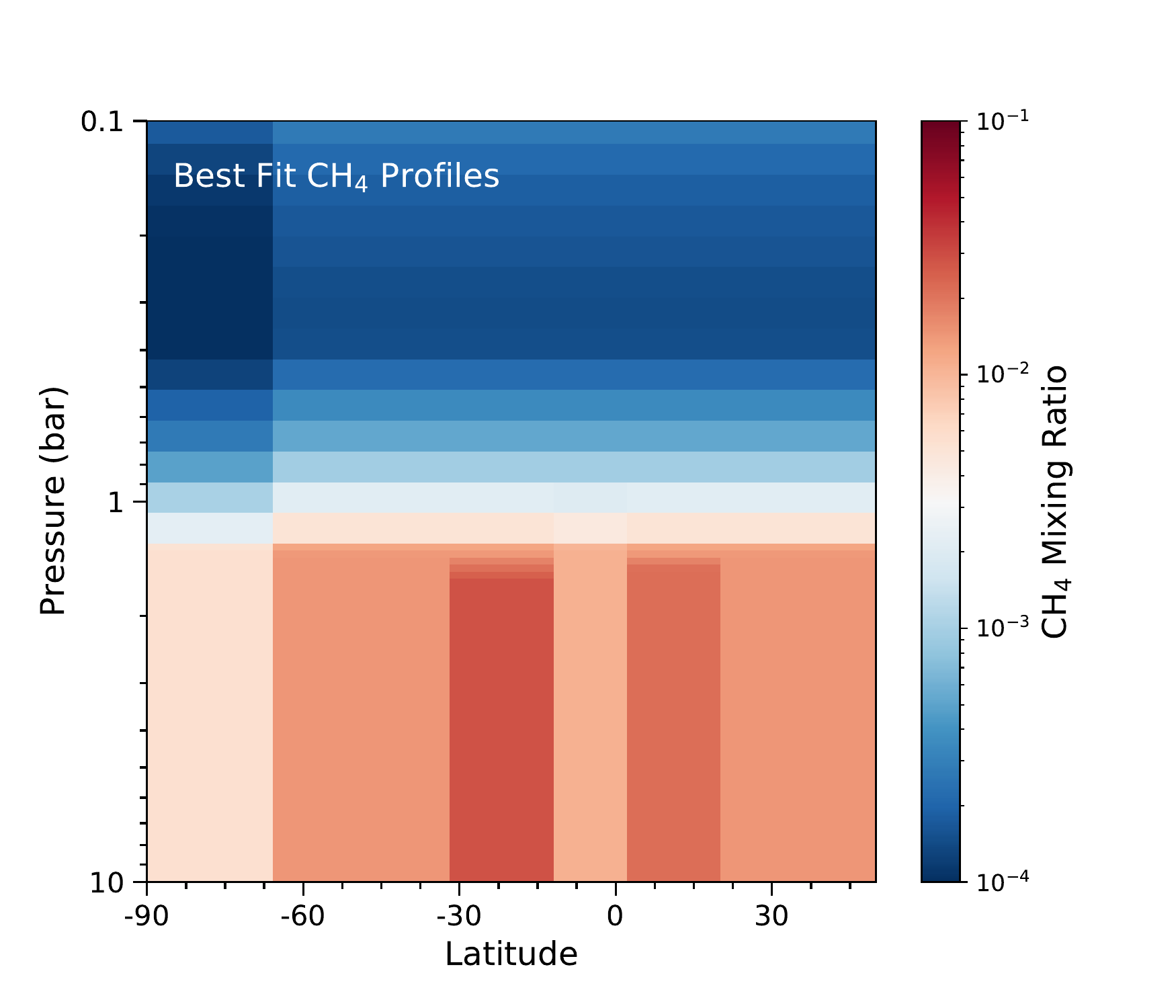}}%
\caption{Left: Best fit H$_2$S vertical profiles versus latitude. Right: Best fit CH$_4$ vertical profiles versus latitude. These profiles are the same as those listed in the first row of Table \ref{table:modeldesc}. The colors represent volume mixing ratios, with high abundances in red and low abundances in blue. Note the different pressure and mixing ratio scales between the two figures.}
\label{fig:best_fit_profs}
\end{figure}

\begin{figure}
\centering
  \includegraphics[width=0.75\linewidth]{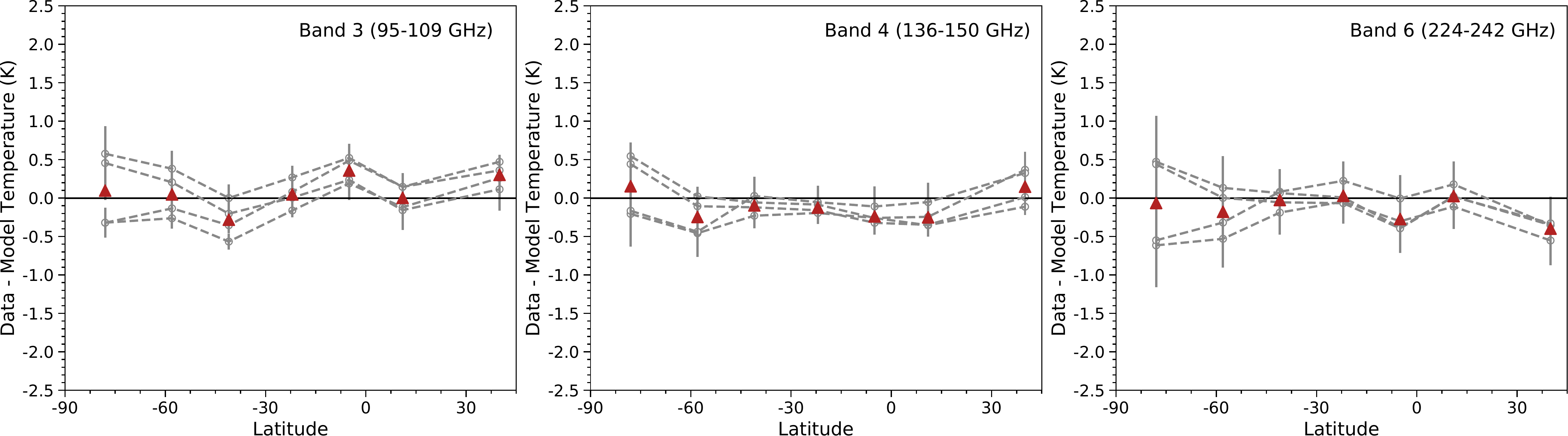}
  \caption{As Figure \ref{fig:residuals_h2s}, but using the best fit H$_2$S and CH$_4$ profiles.}
  \label{fig:residuals_bestfit}
\end{figure}

The first row in Table \ref{table:modeldesc} describes the H$_2$S and CH$_4$ profiles that provide a best fit to the data in all ALMA bands in each latitudinal bin. Figure \ref{fig:best_fit_profs} plots colormaps of the best fit profiles versus latitude. Figure \ref{fig:residuals_bestfit} plots the temperature residuals in each band, showing that an excellent fit is obtained at all latitudes. In general, we find that bright regions in the residual maps require depleting H$_2$S and sometimes CH$_4$ (down to $\sim0.6 - 1.1\%$), while dark regions require supersaturing H$_2$S and enhancing the deep CH$_4$ abundance ($\sim2.2 - 2.9\%$) to fit the data in every band.

\subsubsection{Comparison to \citealp{Karkoschka2011}}

\citet{Karkoschka2011} determined methane mixing ratios from the HST/STIS spectrograph data between $300-1000$ nm across Neptune's disk. Their results are consistent with a constant deep methane mixing ratio of $4\pm1 \%$ at $P > 3.3$ bar. However, between $1.2-3.3$ bar, they argued that the methane mixing ratio was depressed by a factor of $\sim3$ at Neptune's mid-latitudes compared to the equator. Not only is their deep methane mixing ratio significantly higher than our best fit models (ranging between $0.6-2.9\%$), but their observed trend in the methane abundance across Neptune's disk differs from ours. We find high methane mixing ratios relative to the nominal model between $32^{\circ}$S$-12^{\circ}$S and $2^{\circ}$N$-20^{\circ}$N, and low abundances from $90^{\circ}$S$-66^{\circ}$S and $12^{\circ}$S$-2^{\circ}$N. The ALMA Band 6 contribution functions peak at $P<4$ bar, so we expect some signature of the opacity from $1.2-3.3$ bar to appear in our results.

In order to test the models presented in \citet{Karkoschka2011} to our data, we compare how the H$_2$S abundance must change to fit the data in every band. Between $32^{\circ}$S$-20^{\circ}$N, we use the \citet{Karkoschka2011} methane profile at $6^{\circ}$S; elsewhere, we use their $45^{\circ}$S model (see their Figs. 10 and 14)\footnote{The methane profiles in \citet{Karkoschka2011} are constructed by increasing the methane mixing ratio at a constant rate below 1.2 bar: $\sim0.15$ bar/$\%$ at $6^{\circ}$S and $\sim0.6$ bar/$\%$ at 45$^{\circ}$S. There is a transition region using intermediate rates from $20^{\circ}$S$-45^{\circ}$S. A $4\%$ deep mixing ratio is assumed. For $P < 1.2$ bar, methane follows the saturation vapor pressure curve.}. The one exception is at the south pole cap, $90^{\circ}$S$-66^{\circ}$S, where we use our best fitting profile since the deep H$_2$S abundance is fairly well constrained in VLA data. 

\begin{figure}
\centering
  \includegraphics[width=0.75\linewidth]{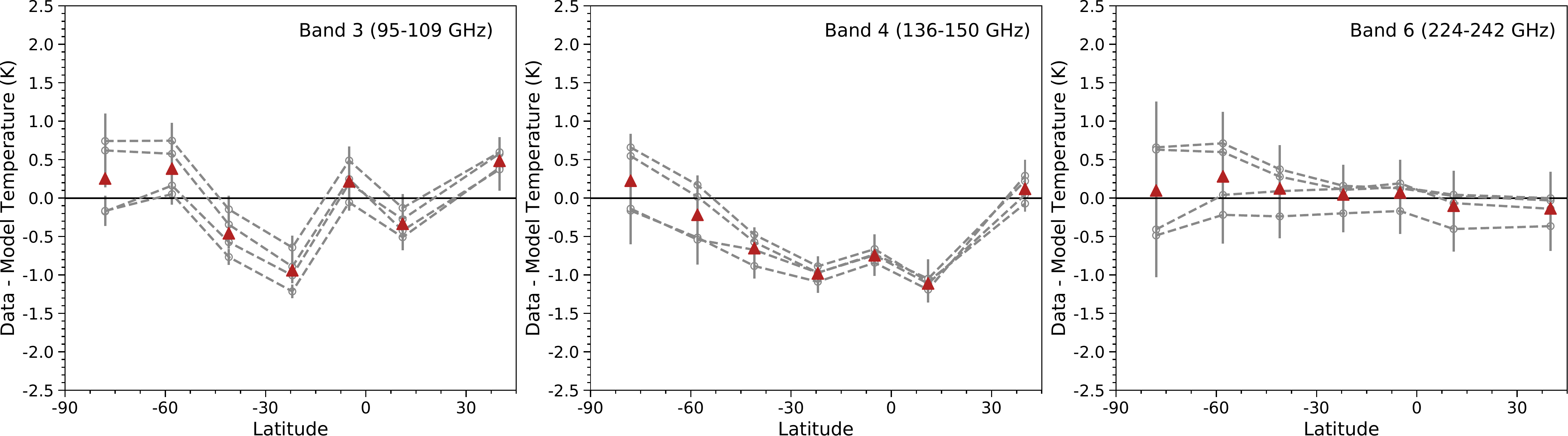}
  \caption{As Figure \ref{fig:residuals_h2s30}, but using the CH$_4$ profiles from \citealp{Karkoschka2011} and finding the H$_2$S profiles that best match Band 6.}
  \label{fig:residuals_kark}
\end{figure}

Since the Band 6 data have the most overlap with the analysis in \citet{Karkoschka2011} in terms of altitude, we find the H$_2$S profiles at each latitude that fit Band 6 well, assuming their adopted CH$_4$ profiles and equilibrium H$_2$. Our results are plotted in Figure \ref{fig:residuals_kark} and a full description of the model and corresponding statistics is listed in Tables \ref{table:modeldesc} and \ref{table:chisq}. We can find H$_2$S profiles which result in generally good agreement between the model and data only from $90^{\circ}$S$-50^{\circ}$S and northward of $20^{\circ}$N. In this case, the disk-averaged H$_2$S abundance is $\sim 10\times$S. \citet{LuszczCook2013a} computed the disk-averaged brightness temperature assuming a 10$\times$S model, showing that the brightness temperature is too high in this case compared to most of the radio data (see also Fig. \ref{fig:diskavgtemp}). Therefore, our data are not consistent with the \citet{Karkoschka2011} methane profiles. We address this in the following section.

\begin{sidewaystable}

\caption{Model names and their constituent profiles in each latitude bin that give a best fit to the data.} 
\begin{tabular}{llllllll}
\hline
\\ \hline
 Model Name & 90$^{\circ}$S$-66^{\circ}$S & 66$^{\circ}$S$-55^{\circ}$S & 55$^{\circ}$S$-32^{\circ}$S & 32$^{\circ}$S$-12^{\circ}$S & 12$^{\circ}$S$-2^{\circ}$N & 2$^{\circ}$N$-20^{\circ}$N & 20$^{\circ}$N$-50^{\circ}$N \\ \hline
 Best Fit & \begin{tabular}{@{}l@{}}H$_2$S 1.5$\times$S \\ CH$_4$ 0.55$\%$ \\ H$_2$ equil.\end{tabular} & \begin{tabular}{@{}l@{}}H$_2$S 10$\times$S \\ CH$_4$ 1.44$\%$ \\ H$_2$ equil.\end{tabular} & \begin{tabular}{@{}l@{}}H$_2$S 30$\times$S \\ CH$_4$ 1.44$\%$ \\ H$_2$ equil.\end{tabular} & \begin{tabular}{@{}l@{}}H$_2$S supersat-v1 \\ CH$_4$ 2.88$\%$ \\ H$_2$ equil.\end{tabular} & \begin{tabular}{@{}l@{}}H$_2$S 20$\times$S \\ CH$_4$ 1.08$\%$ \\ H$_2$ equil.\end{tabular} & \begin{tabular}{@{}l@{}}H$_2$S supersat-v1 \\ CH$_4$ 2.16$\%$ \\ H$_2$ equil.\end{tabular} & \begin{tabular}{@{}l@{}}H$_2$S 10$\times$S \\ CH$_4$ 1.44$\%$ \\ H$_2$ equil.\end{tabular} \\ \hline
 Vary H$_2$S Only & \begin{tabular}{@{}l@{}}H$_2$S 1.5$\times$S \\ CH$_4$ 0.55$\%$ \\ H$_2$ equil.\end{tabular} & \begin{tabular}{@{}l@{}}H$_2$S 10$\times$S \\ CH$_4$ 1.44$\%$ \\ H$_2$ equil.\end{tabular} & \begin{tabular}{@{}l@{}}H$_2$S 30$\times$S \\ CH$_4$ 1.44$\%$ \\ H$_2$ equil.\end{tabular}  & \begin{tabular}{@{}l@{}}H$_2$S supersat-v2\\ CH$_4$ 1.44$\%$ \\ H$_2$ equil.\end{tabular}  &  \begin{tabular}{@{}l@{}}H$_2$S 3$\times$S \\ CH$_4$ 1.44$\%$ \\ H$_2$ equil.\end{tabular}  &  \begin{tabular}{@{}l@{}}H$_2$S supersat-v2 \\ CH$_4$ 1.44$\%$ \\ H$_2$ equil.\end{tabular}  &  \begin{tabular}{@{}l@{}}H$_2$S 3$\times$S \\ CH$_4$ 1.44$\%$ \\ H$_2$ equil.\end{tabular} \\ \hline
 Vary CH$_4$ Only &  \begin{tabular}{@{}l@{}}H$_2$S 30$\times$S \\ CH$_4$ 0.55$\%$ \\ H$_2$ equil.\end{tabular} & \begin{tabular}{@{}l@{}}H$_2$S 30$\times$S \\ CH$_4$ 0.55$\%$ \\ H$_2$ equil.\end{tabular} & \begin{tabular}{@{}l@{}}H$_2$S 30$\times$S \\ CH$_4$ 1.44$\%$ \\ H$_2$ equil.\end{tabular} & \begin{tabular}{@{}l@{}}H$_2$S 30$\times$S \\ CH$_4$ 2.88$\%$ \\ H$_2$ equil.\end{tabular} & \begin{tabular}{@{}l@{}}H$_2$S 30$\times$S \\ CH$_4$ 1.44$\%$ \\ H$_2$ equil.\end{tabular} & \begin{tabular}{@{}l@{}}H$_2$S 30$\times$S \\ CH$_4$ 1.44$\%$ \\ H$_2$ equil.\end{tabular} & \begin{tabular}{@{}l@{}}H$_2$S 30$\times$S \\ CH$_4$ 0.55$\%$ \\ H$_2$ equil.\end{tabular} \\ \hline
 Vary H$_2$ Only &  \begin{tabular}{@{}l@{}}H$_2$S 30$\times$S \\ CH$_4$ 0.55$\%$ \\ H$_2$ equil.\end{tabular} & \begin{tabular}{@{}l@{}}H$_2$S 30$\times$S \\ CH$_4$ 1.44$\%$ \\ H$_2$ equil.\end{tabular} & \begin{tabular}{@{}l@{}}H$_2$S 30$\times$S \\ CH$_4$ 1.44$\%$ \\ H$_2$ equil.\end{tabular} & \begin{tabular}{@{}l@{}}H$_2$S 30$\times$S \\ CH$_4$ 1.44$\%$ \\ H$_2$ 90$\%$ equil.\end{tabular} & \begin{tabular}{@{}l@{}}H$_2$S 30$\times$S \\ CH$_4$ 1.44$\%$ \\ H$_2$  equil.\end{tabular} & \begin{tabular}{@{}l@{}}H$_2$S 30$\times$S \\ CH$_4$ 1.44$\%$ \\ H$_2$ equil.\end{tabular} & \begin{tabular}{@{}l@{}}H$_2$S 30$\times$S \\ CH$_4$ 1.44$\%$ \\ H$_2$ equil.\end{tabular} \\ \hline
  KT2011$^a$ CH$_4$ &  \begin{tabular}{@{}l@{}}H$_2$S 1.5$\times$S \\ CH$_4$ 0.55$\%$ \\ H$_2$ equil.\end{tabular} & \begin{tabular}{@{}l@{}}H$_2$S 5$\times$S \\ CH$_4$ 45S \\ H$_2$ equil.\end{tabular} & \begin{tabular}{@{}l@{}}H$_2$S 15$\times$S \\ CH$_4$ 45S \\ H$_2$ equil.\end{tabular} & \begin{tabular}{@{}l@{}}H$_2$S $15\times$S\\ CH$_4$ 6S \\ H$_2$ equil.\end{tabular} & \begin{tabular}{@{}l@{}}H$_2$S 10$\times$S \\ CH$_4$ 6S \\ H$_2$  equil.\end{tabular} & \begin{tabular}{@{}l@{}}H$_2$S 10$\times$S \\ CH$_4$ 6S \\ H$_2$ equil.\end{tabular} & \begin{tabular}{@{}l@{}}H$_2$S 5$\times$S \\ CH$_4$ 45S\\ H$_2$ equil.\end{tabular} \\ \hline
\end{tabular}

\footnotesize{$^a$ KT2011 refers to \citealp{Karkoschka2011}. CH$_4$ 45S and 6S refer to their adopted methane profiles at those latitudes in their Figs. 10 and 14}  \\
\label{table:modeldesc}
\end{sidewaystable}

\begin{table}

\begin{tabular}{lccccccc}
 \hline
\\ \hline
 Model Name & 90$^{\circ}$S-66$^{\circ}$S & 66$^{\circ}$S-55$^{\circ}$S & 55$^{\circ}$S-32$^{\circ}$S & 32$^{\circ}$S-12$^{\circ}$S & 12$^{\circ}$S-2$^{\circ}$N & 2$^{\circ}$N-20$^{\circ}$N & 20$^{\circ}$N-50$^{\circ}$N \\ \hline
 Best Fit & 3.4 & 2.9 & 6.8 & 2.0 & 8.1 & 4.1 & 6.2 \\
 Vary H$_2$S Only & 3.4 & 2.6 & 3.2 & 14.8 & \textbf{46.4} & 13.0 & 7.5  \\
 Vary CH$_4$ Only & \textbf{55.8} & \textbf{20.7} & \textbf{16.5} & \textbf{56.7} & 2.3 & \textbf{25.5} & 7.3 \\
 Vary H$_2$ Only & \textbf{95.8} & \textbf{48.0} & \textbf{16.1} & \textbf{85.6} & 2.8 & \textbf{17.2} & \textbf{29.0}\\
 KT2011 CH$_4$ Only & 3.9 & 4.3 & \textbf{28.9} & \textbf{100.4} & \textbf{20.9} & \textbf{50.3} & 8.8\\ \hline
\end{tabular}

\caption{Reduced $\chi^2$ values versus latitude bin per model. Bold values are significant at the $p \leq 0.05$ level, meaning the model at those latitudes is inconsistent with the observations.} 
\label{table:chisq}
\end{table}

\section{Discussion}
\label{S:6}

\subsection{Neptune's South Polar Cap}
Our ALMA millimeter observations provide a glimpse of Neptune's atmosphere situated  between the deep troposphere in centimeter maps ($P \geq 10$ bar) and the upper troposphere and stratosphere in the visible and infrared. The constraints on the trace gases are useful to infer the dynamics of Neptune's atmosphere. Of particular interest is Neptune's south pole, whose high temperatures were first published by \citet{Hammel2007b} with images taken with the Gemini north telescope at 7.7$\mu$m and 11.7$\mu$m. They suggested these bright regions are due to enhancements in ethane and methane. \citet{Orton2007} imaged atmospheric line-free thermal emission of Neptune with the Very Large Telescope in 2006, finding temperature excesses of $10-11$K and $3-5$K near, but not at, Neptune's south pole at 17.6$\mu$m and 18.7$\mu$m respectively. These authors suggested that seasonal warming around Neptune's south pole could explain why the stratospheric abundance of methane is larger than expected; cold temperatures should result in methane condensing and becoming trapped below the tropopause. However, warm polar temperatures may allow methane gas to escape upward into the stratosphere and diffuse across the globe.

 Warm brightness temperature measurements at high-latitudes are persistent throughout radio maps. \citet{dePater2014} found temperature enhancements from 8$-$30K in VLA 1.3$-$6.2 cm maps, where sensitivities peak between $5-50$ bar. EVLA 1-cm maps show enhancements of similar magnitudes southward of 70$^{\circ}$S \citep{Butler2012DPS}. \citet{LuszczCook2013b} see southern high-latitude enhancements of $2-3$K compared to the northern mid-latitudes in 1.2 mm CARMA maps. Our ALMA maps show average enhancements of $2-3$K in Bands 3 and 4, with sensitivities peaking at $P > 1$ bar, and $1-2$K in Band 6, whose sensitivities peak at 1 bar. \citet{Iino2018} analyzed ALMA flux calibration data of Neptune at 646 GHz (0.46 mm), peaking at 0.6 bar, and ruled out a detection of the south polar hot spot greater than 2.1K compared to the background. Combined, these data suggest that the magnitude of the south polar brightness enhancements decreases with increasing altitude. This trend is likely due to the temperature-pressure profile, which is mostly isothermal between $0.1-1$ bar. This picture also appears dynamically distinct from that described in \citet{Orton2007}, who predicted upwelling air to explain methane and ethane enrichment in the stratosphere. In the radio, brightness enhancements are consistent with low-opacity (dry) air so deeper warmer layers are probed. The air is likely subsiding, after having been dried out at other latitudes. The subsiding air causes adiabatic warming, which is sensed in the mid-IR (e.g., Fig. 16 in \citet{dePater2014}). Persistent cloud activity surrounding Neptune's south pole may be indicative of vigorous convection and evidence of a south polar vortex \citep{LuszczCook2010}, analogous to the polar activity seen on Saturn \citep{Fletcher2008, Dyudina2008}. Such a system could explain the observed temperature, ethane and methane enhancements in the mid-IR, polar cloud features seen with \textit{Voyager} and Keck, and high brightness temperatures in the radio.
 
Recent findings by \citet{Irwin2019} find a tentative detection of an H$_2$S spectral signature between $1.57-1.58\mu$m on Neptune with Gemini-North/NIRS. They find that the signature is stronger at southern latitudes than at the equator, with H$_2$S abundances around 3 ppm and 1 ppm respectively at the top of Neptune's H$_2$S cloud deck: $2.5-3.5$ bar. This broadly agrees with our suite of proposed H$_2$S profiles (Fig. 9), where H$_2$S is depleted to an abundance of 1 ppm between $\sim 2.5-4$ bar. The exception is our model for the south polar cap, which has less H$_2$S at the relevant altitudes than predicted by \citet{Irwin2019}. In order to match their results, we would have to lower the altitude where H$_2$S follows the saturated vapor pressure curve (to $P > 4.5$ bar) so that the abundance from $2.5-3.5$ bar could be increased to 1 ppm. Moreover, their retrieved H$_2$S abundances are most consistent with models which deplete the deep abundance of CH$_4$ at the southern high-latitudes relative to the equator. This agrees with our results and \citet{Karkoschka2011}. 
  
\subsection{Neptune's Mid-Latitudes and Equator}

Moving northward, our ALMA maps show that latitudes spanning $32^{\circ}-12^{\circ}$S and $2^{\circ}-20^{\circ}$N are colder than the background by $\sim0.5-1.5$K in all three bands. This difference can only be explained at all wavelengths with models which both increase the deep CH$_4$ abundance to $\sim2.2-2.9\%$ and supersaturate the H$_2$S profile. This is consistent with upwelling, adiabatically cooling plumes, and the observed distribution of prevalent bright cloud activity in this latitude range seen in the visible and near-IR. \citet{Karkoschka2011} find that methane is well-mixed in the deep-atmosphere, with an abundance of $4\pm1\%$. In addition, they find that methane is depleted by a factor of 3 at the southern mid-latitudes compared to the equatorial region between 1.2 and 3.3 bar. This pressure range contributes the most to the opacity in Band 6, wavelengths where we also see the lowest temperature contrasts. If we assume their horizontal methane profiles, our models are too cold to match the ALMA data, due to their high methane abundance. A corresponding global decrease in the H$_2$S abundance to $\sim10\times$S must be made in order to decrease the total opacity and increase the brightness temperature so that the Band 6 data and model agree. However, decreasing the global H$_2$S abundance has the side effect of dramatically increasing the brightness temperatures in Bands 3 and 4, particularly near the equator. In Band 4, these profiles fit the equatorial regions poorly, yet agree at these latitudes in Band 3. For Band 4 to match here, the H$_2$S abundance would have to be increased to account for a 1 K difference. However, the Band 3 model temperatures would be decreased by at least this same amount, putting it outside a two-sigma fit to the data. Moreover, model disk-averaged brightness temperatures assuming a $10\times$ S H$_2$S abundances are too warm compared to the ALMA and 2003 VLA data (Fig. \ref{fig:diskavgtemp}).

Our results are more consistent with \citet{Baines1995}, who predict disk-averaged deep-atmosphere methane molar fractions of $2.2^{+0.5}_{-0.6}\%$. \citet{Karkoschka2011} remark that the two studies use different haze and cloud profiles and relative humidities. Correcting for these factors lowers their deep-atmosphere methane mixing ratio to within the error bars cited in \citet{Baines1995}. In addition, \citet{Karkoschka2011} note that a significant contribution to their error bars is due to systematic variations in their center-to-limb profiles in the methane bands. Their model is too bright at disk center and too cold at the limb. Lowering the methane mixing ratio and using different values for the (poorly known) methane and hydrogen absorption coefficients would improve their fit. Alternatively, a high deep methane mixing ratio may be possible if methane is subsaturated, which we did not consider in our models. Subsaturated models will produce warmer brightness temperatures while increasing the deep methane mixing ratio will produce colder temperatures. As a result, these effects will partially cancel and combined may produce a more reasonable fit. There may also be uncertainties in the millimeter spectral line parameters that are not accounted for here. Laboratory measurements of H$_2$S absorption in the millimeter are extremely limited and not at Neptune's tropospheric conditions \citep{{Joiner1992}}. Updated measurements of the trace gas absorption lines under cold, ice giant conditions would improve, and perhaps reconcile, models of these multi-wavelength observations.

While we disagree with the deep methane mixing ratio computed by \citet{Karkoschka2011}, we find general agreement with their observed latitude trends. Both our study and theirs find enhancements in trace gases from $32^{\circ}-12^{\circ}$S and $2^{\circ}-20^{\circ}$N, depletion from $90^{\circ}-50^{\circ}$S and north of 20$^{\circ}$N, and intermediate values from $50^{\circ}-32^{\circ}$S. Our sole inconsistency is from $12^{\circ}$S$-2^{\circ}$N, where we also depleted trace gases. However, at these latitudes, \citet{Karkoschka2011} lowered the tropospheric haze optical depth compared to the surrounding latitudes. If their haze optical depth was increased here, we would expect them to need to decrease the methane abundance from $12^{\circ}$S$-2^{\circ}$N to fit their data. \citet{LuszczCook2016} also investigated methane profiles in cloud free regions of Neptune with the OSIRIS integral field spectrograph in the H and K broad bands. These data probe altitudes higher than ALMA and HST/STIS and so are insensitive to the deep troposphere methane abundance. These authors saw tentative evidence of meridional variations in the methane profile, qualitatively consistent with \citet{Karkoschka2011}, but remarked that their parameterizations could not fully characterize the true shape of the methane profile since they were only sensitive to $P < 2.5$ bar.

\begin{figure}
\centering
  \includegraphics[width=0.75\linewidth]{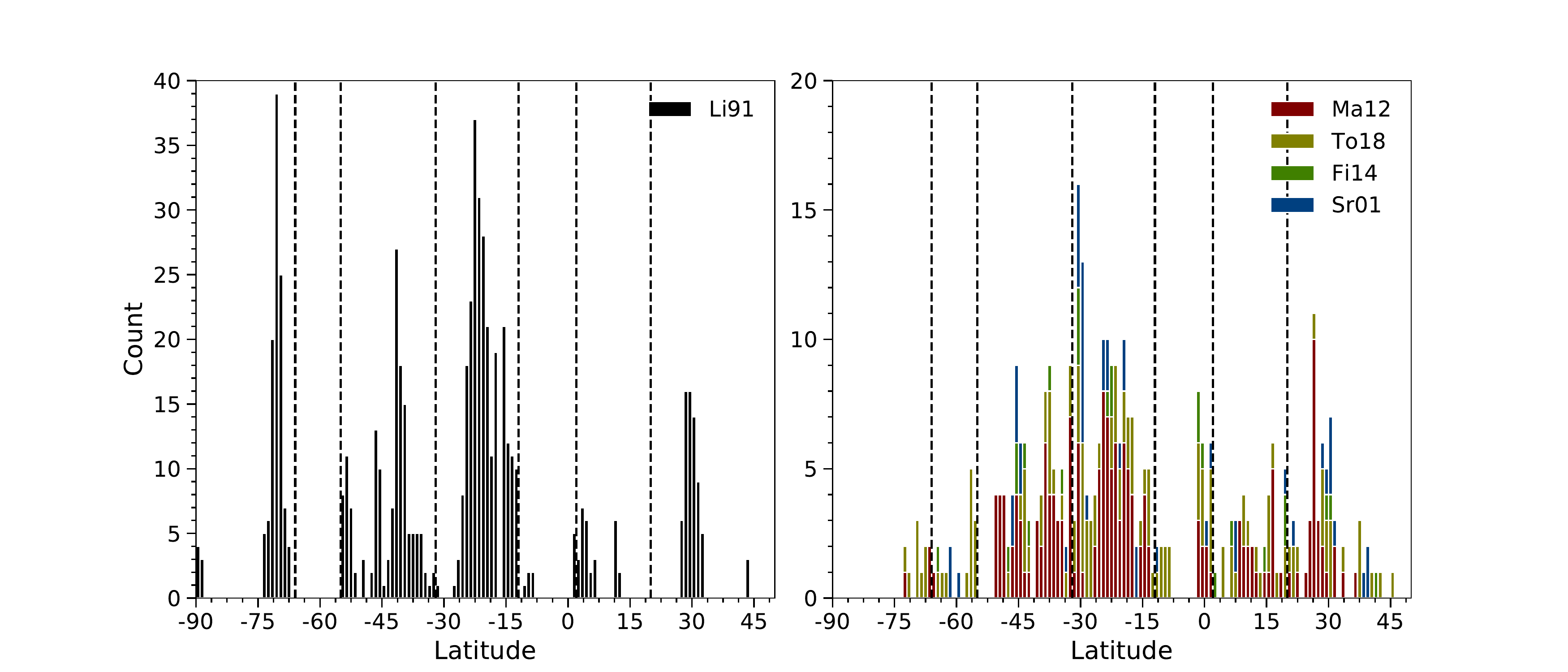}
  \caption{Histogram counting the number of tracked bright cloud features versus latitude. Dashed lines delineate the latitude bands used in the modeling of the ALMA data and correspond to variations in the residual brightness temperature. Counts are in one-degree latitude bins and latitudes have been converted to planetographic coordinates. Since the \textit{Voyager} data would otherwise dominate the count and perceived trend, these data are split apart, showing the \textit{Voyager} counts on the left and post-\textit{Voyager} counts on the right. These data come from five papers, labeled as follows: Li91 - \citealp{Limaye1991}; Ma12 - \citealp{Martin2012}; To18 - \citealp{Tollefson2018}; Fi14 - \citealp{Fitzpatrick2013}; Sr01 - \citealp{Sromovsky2001b}.}
  \label{fig:featurehist}
\end{figure}

In the equatorial region, our ALMA residual maps show clear latitudinal structure undetected in prior radio studies of Neptune. North of the equator, we detect low brightness temperatures that are consistent with supersaturating H$_2$S and a slight increase in the deep CH$_4$ abundance. In the visible and near-IR, Neptune's equatorial region is quiescent, lacking bright cloud activity compared to the dramatic stormy mid-latitudes. However, there is evidence of more cloudy activity just north of the equator than south of it, agreeing with our finding of CH$_4$ enrichment from $2^{\circ}-20^{\circ}$N and depletion from $12^{\circ}$S$-2^{\circ}$N. Figure \ref{fig:featurehist} shows a histogram counting the number of bright cloud features versus latitude from \textit{Voyager}, the Hubble Space Telescope, and H-band Keck maps which were tracked in five papers: \citet{Limaye1991, Sromovsky2001b, Martin2012, Fitzpatrick2013, Tollefson2018}. By eye, there appears to be a clear correlation between our ALMA defined latitude bands and latitudes where the number of features transitions from low-to-high or vice-versa, particularly in the \textit{Voyager} data. This suggests that the banded structure seen by ALMA from $1-10$ bar exists up to the visible cloud deck. Specifically, at the central latitudes within the 12$^{\circ}$S$-$2$^{\circ}$S band, there is a persistent scarcity of activity observed over a 30 year period. In contrast, features are seen regularly between 2$^{\circ}$S$-$20$^{\circ}$N, suggesting a larger source of condensible methane. This agrees with our best fit model which depletes CH$_4$ to $\sim1.1\%$ in the cloud-free latitudes and enriches it to $\sim2.2\%$ in the cloud-prevalent latitudes. The enhancement of methane in this region is consistent with a moist convective origin of a recently discovered large bright storm centered at 2$^{\circ}$N \citep{Molter2019}. Moreover, this model may explain inconsistencies between the thermal wind equation and observed vertical wind shear measurements. \citet{Fitzpatrick2013} and \citet{Tollefson2018} detected vertical wind shear at Neptune's equator by tracking bright cloud features in the H- and K'-bands with Keck, with the K'-band probing higher altitudes with features that have, on average, stronger retrograde velocities than features seen in the H-band. \citet{Tollefson2018} showed that the equator must be methane-rich and warm compared to mid-latitudes if the thermal wind equation holds.

\begin{figure}
\centering
  \includegraphics[width=0.75\linewidth]{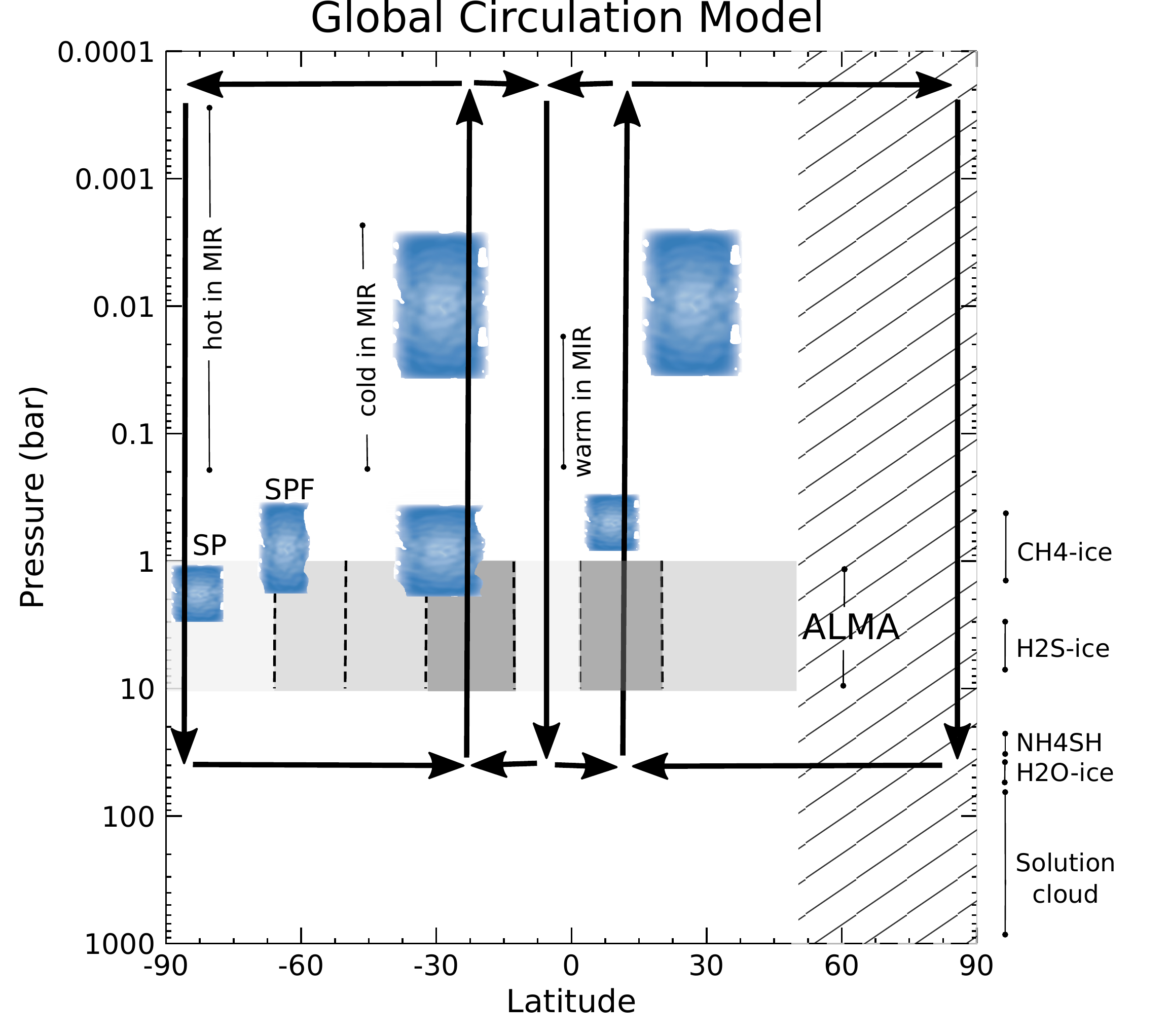}
  \caption{Schematic of Neptune's global circulation inferred from ALMA observations, adapted from Fig. 21 in \citet{dePater2014}. Black arrows outline the circulation pattern, which extends from the stratosphere down to 40 bar. The biggest change from the sketch in \citet{dePater2014} is our circulation cells are narrowed near the equator and we prefer a subsiding region which is just south of the equator (aligning with the $12^{\circ}$S$-2^{\circ}$N range seen in our residual map), instead of hemispheric symmetric cells. The gray rectangle between $1-10$ bar shows the sensitivity of the ALMA spectral bands (see Fig. 1). Dark-gray (off-white) rectangles are the latitude ranges where the residual brightness temperature is colder (warmer) than the background (see Fig. 4). At these latitudes, both H$_2$S and CH$_4$ are enriched (depleted) (see Table 5). The light-gray rectangles are latitude ranges where the residual brightness temperature is similar to the background. Here, H$_2$S and CH$_4$ equal their nominal value, or only H$_2$S is depleted. The locations of Visible/NIR clouds are illustrated with blue vertical patches. Most clouds are seen at the mid-latitudes (peak cloud counts are centered near $25^{\circ}$S and $30^{\circ}$N, see Fig. 15) and are seen high in the atmosphere (e.g., those tracked in \citet{dePater2014}). However, clouds just north of the equator are also seen and are deeper in the atmosphere (\citet{dePater2014},  \citet{Tollefson2018}, and Fig. 15). Neptune's south polar feature (SPF) and south polar (SP) cloud are also placed. Marked cloud layers (CH$_4$-ice, H$_2$S-ice, NH$_4$SH, H$_2$O-ice, Solution Cloud) on the right are assumed to be independent of latitude (see Fig. 1). Also indicated are where high and low temperatures are measured in the MIR (Fletcher et al. 2014). Poleward of $\sim50^{\circ}$N, seeing is cut off in all observations as Neptune's north pole is tilted away from the observer (rectangular diagonal hatches). Downwelling motions are assumed at Neptune's north pole in order to complete the circulation diagram.}
  \label{fig:circulation}
\end{figure}

The thermal profile inferred from MIR observations, the horizontal distribution of clouds, and observed regions of enriched/depleted air all relate to vertical circulation motions. Based on multi-wavelength observations, \citet{dePater2014} suggest a vertically extended, hemispheric symmetric double-cell pattern with upwelling at the mid-latitudes and downwelling at the equator and poles from the stratosphere down to the deep troposphere ($\sim$40 bar). The ALMA, HST/STIS, and OSIRIS results suggest more detailed circulation, particularly near the equator. Our analysis is consistent with the prediction of vertically extended cells, going from the stratosphere down to $\sim40$ bar, since there is a clear alignment between the ALMA detected latitudinal bands and the observed distribution of cloud features. However, we argue that the equatorial region is more intricate than what was outlined by \citet{dePater2014}, with upwelling from 2$^{\circ}$N$-$20$^{\circ}$N and downwelling from 12$^{\circ}$S-2$^{\circ}$N. A narrower circulation cell than predicted by \citet{dePater2014} centered just south of the equator would partially explain the differing predictions between these studies. Figure \ref{fig:circulation} gives a sketch of our proposed global circulation pattern.

\section{Conclusion}
\label{S:7}
Spatially resolved millimeter maps of Neptune are presented in three ALMA bands, spanning $95-242$ GHz. These maps have unprecedented sensitivities in the millimeter, ranging between $0.1-0.6$K, and resolutions down to $0.3"$, roughly one-eighth of Neptune's diameter. The observed emission is mainly modulated by the opacity due to H$_2$S absorption, CIA of H$_2$ with H$_2$, He, and CH$_4$, and \textit{ortho/para} H$_2$. We used the radiative transfer code Radio-BEAR to vary the abundance profiles of these gases in order to model the brightness temperature across Neptune's disk. Our main conclusions are as follows:

\begin{enumerate}
    \item The disk-averaged brightness temperature of Neptune in the millimeter and centimeter matches a model where: the temperature-pressure profile follows a dry adiabat; H$_2$S, CH$_4$, and H$_2$O are enriched by 30$\times$ their protosolar values, while NH$_3$ is held at $1\times$ its protosolar value; intermediate H$_2$ is assumed. This is referred to as the `nominal' model.
    \item Subtracting the nominal beam-convolved model from the data shows clear brightness temperature variations across Neptune's disk. We identify latitudes in between significant transitions in the brightness temperature:  90$^{\circ}-66^{\circ}$S, 66$^{\circ}-55^{\circ}$S, 55$^{\circ}-32^{\circ}$S, 32$^{\circ}-12^{\circ}$S, 12$^{\circ}$S$-$2$^{\circ}$N, 2$^{\circ}-20^{\circ}$N, and northward of 20$^{\circ}$N. These bands are at least the size of the ALMA synthesized beam and are apparent in all maps. Relative to the nominal model, brightness enhancements of $1-3$K are seen at $90^{\circ}$S$-$$66^{\circ}$S and $66^{\circ}$S$-$$55^{\circ}$S. Negative temperature residuals between $0.5-1.5$K are seen from 32$^{\circ}-12^{\circ}$S and 2$^{\circ}-20^{\circ}$N. These bands align with regions transitioning from high or low counts in the number of cloud features versus latitude, suggesting that the banded structure we see in the ALMA data exists up at the visible cloud deck. As a result, these identified latitudinal bands may be indicative of a zonal wind profile that is more complex than hitherto considered.
    \item At the south polar cap, our best fit model depletes the deep atmospheric abundance of both H$_2$S to 1.5$\times$ the protosolar value and CH$_4$ to 0.55$\%$ (11.5$\times$S). This is consistent with models fitting VLA and CARMA data \citep{dePater2014, LuszczCook2013b}. Between 55$^{\circ}-32^{\circ}$S and northward of 20$^{\circ}$N, our best fit models deplete H$_2$S to $10\times$S while keeping CH$_4$ at the nominal value of 1.44$\%$ (30$\times$S). From  32$^{\circ}-12^{\circ}$S and 2$^{\circ}$N$-$20$^{\circ}$N, H$_2$S is supersaturated and the deep abundance of CH$_4$ is enriched to $2.2-2.9\%$ ($45-60\times$S). From 12$^{\circ}$S$-2^{\circ}$N, we deplete H$_2$S to 10$\times$ protosolar and CH$_4$ to 1.1$\%$ ($22.5\times$S). Warm brightness temperatures relative to the nominal model are consistent with dry, subsiding air. Conversely, cold brightness temperatures are consistent with moist, rising air. Our results are, therefore, consistent with an intricate global circulation system that extends from the cloud deck to deep in the atmosphere.
    
\end{enumerate}

These ALMA maps are evidence of a more complex zone and belt structure that may be yet unresolved in the zonal wind field. An ice-giant probe and high-resolution spacecraft imaging characterizing the zonal wind structure in detail would help settle any inconsistencies between Neptune's velocity, thermal, and compositional profiles.

\section{Acknowledgements}

The authors wish to thank Arielle Moullet for helping with the data reduction and imaging process. This work has been supported in part by the National Science Foundation, NSF Grant AST-1615004 to UC Berkeley and by NASA Headquarters: under the NASA Earth and Space Science Fellowship program Grant NNX16AP12H to UC Berkeley. 

This paper makes use of the following ALMA data: ADS/JAO.2016.1.00859.S. ALMA is a partnership of ESO (representing its member states), NSF (USA) and NINS (Japan), together with NRC (Canada), NSC and ASIAA (Taiwan), and KASI (Republic of Korea), in cooperation with the Republic of Chile. The Joint ALMA Observatory is operated by ESO, AUI/NRAO and NAOJ. The National Radio Astronomy Observatory	is a facility of the National Science Foundation operated under cooperative agreement	by Associated Universities, Inc.

\end{document}